**The entrepreneurial logic of startup software development: A study of 40 software startups**

Anh Nguyen-Duc[1] · Kai-Kristian Kemell[2] · Pekka Abrahamsson[3]

**Abstract**

Context: Software startups are an essential source of innovation and software-intensive products. The need to understand product development in startups and to provide relevant support are highlighted in software research. While state-of-the-art literature reveals how startups develop their software, the reasons why they adopt these activities are underexplored.

Objective: This study investigates the tactics behind software engineering (SE) activities by analyzing key engineering events during startup journeys. We explore how entrepreneurial mindsets may be associated with SE knowledge areas and with each startup case.

Method: Our theoretical foundation is based on causation and effectuation models. We conducted semi-structured interviews with 40 software startups. We used two-round open coding and thematic analysis to describe and identify entrepreneurial software development patterns. Additionally, we calculated an effectuation index for each startup case.

Results: We identified 621 events merged into 32 codes of entrepreneurial logic in SE from the sample. We found a systemic occurrence of the logic in all areas of SE activities. Minimum Viable Product (MVP), Technical Debt (TD), and Customer Involvement (CI) tend to be associated with effectual logic, while testing activities at different levels are associated with causal logic. The effectuation index revealed that startups are either effectuation-driven or mixed-logics-driven.

Conclusions: Software startups fall into two types that differentiate between how traditional SE approaches may apply to them. Effectuation seems the most relevant and essential model for explaining and developing suitable SE practices for software startups.

Keywords: Software startup engineering, entrepreneurial logics, effectuation theory, case study, effectuation index, software engineering for startups

## 1. Introduction

More and more software is developed by startup companies with limited resources and little operating history. Successful companies like Uber, Spotify, and Kahoot developed their software products during their startup stages. According to Pitchbook, investment in US startups only is more than 120 billion USD in 2019 (PitchBook, 2019). This substantial financial investment also implies a massive waste due to startups' high failure rate (Giardino et al., 2014). Previous research reveals critical challenges in both business and product development (Giardino et al., 2015). Consequently, attempts to deal with these challenges could eventually increase the odds of success, and the economic savings would be significant (Lindgren and Münch, 2016). The need to better understand software engineering (SE) in startups and provide relevant support for practitioners has been emphasized in the software startup research community (Unterkalmsteiner, 2016; Pantiuchina, 2017; Bajwa et al., 2017; Nguyen-Duc et al., 2020). The emergence of software startup as a research theme is shown by an increasing number of studies on different engineering aspects in a startup context, for example, SE (Klotins, Unterkalmsteiner, Chatzipetrou, et al., 2019), requirements engineering (Melegati et al., 2019), software architecture (Fagerholm et al., 2017), software Minimum Viable Product (MVP) (Duc and Abrahamsson, 2016), and startup ecosystems (Tripathi et al., 2018). These studies explore the commonalities among startups regarding engineering processes, practices, and ways of working. We have better understood the demand for SE principles, processes, and practices in startup companies, their challenges, and common ways of working. However, we do not understand why they adopt a particular workflow and under which circumstances they make these decisions.

[1] Business School, University of South Eastern Norway
Gullbringvegen 36, 3800, Bø i Telemark
Tel.: +47-48348496
E-mail: angu@usn.no

[2] University of Jyväskylä, Finland
[3] University of Jyväskylä, Finland

State-of-the-art software startup research inherited from empirical SE research several preoccupations with normative studies on methods, methodologies, and models, and it lacks theories to understand and explain how and why things happen (Ralph, 2016). For instance, Aurum et al. (2003) adopted decision-making theories to understand the nature of requirement engineering activities. In response to this theoretical gap in software startup research, our previous work began to explore decision-making logics in software startups (Nguyen-Duc, Seppanen, and Abrahamsson, 2015; Kemell, Ventilä, Kettunen, and Mikkonen, 2019). Understanding the logic behind startup activities would enable the exploration of a systematic connection between decisions, activities, behaviors, and startup context, contributing to theory building in software startup research. Furthermore, patterns, or anti-patterns with their antecedent and consequent factors can be directly beneficial for startup companies.

Startups differ from established companies in the strong presence of entrepreneurial personalities, behaviors, decision-making, and leadership (Bygrave et al., 1991). Startups operate with a high level of uncertainty, multiple influences, and small team sizes, which magnify the influence of key persons, such as the CEO or CTO, on the project's success (Paternoster et al., 2014; Berg et al., 2018; Giardino et al., 2014). While entrepreneurial characteristics are evident in both information systems and business literature (Ojala 2015, 2016; Nambisan 2017), entrepreneurship rarely appears in SE research, either contextually or as a primary focus of the investigation. Tripathi et al. (2018) found that entrepreneurs' backgrounds influence how MVPs are developed. The following year, Melegati et al. (2019) found that startup founders strongly influence requirement engineering activities. However, neither study explores the logic underlying observed phenomena. Prescriptive methodologies have recently attracted considerable interest in entrepreneurship research (Sarasvathy and Dew 2005a, 2005b; Dew et al., 2009; Fisher, 2012; Berends et al., 2013; Reymen et al., 2015; Mansoori and Lackéus, 2019). There is a widespread research effort to identify the common logic or principles behind entrepreneurs' decisions and actions. A prominent example of an entrepreneurial logic is effectuation, presented as a set of heuristics any entrepreneur could use for business development in the context of high uncertainty (Sarasvathy and Dew 2005a, 2005b). The logic has been proposed in contrast to a traditional causation logic, in which entrepreneurs are plan-driven, perform their best within given constraints, and accept the possibility of a changed goal (Sarasvathy and Dew 2005a; Wiltbank et al., 2006; Read et al., 2009). As product development is critical for software startups, it is crucial to understand how entrepreneurial logic applies to software development activities.

In the quest to develop a theory of software startup engineering (Nguven-Duc et al., 2020), we want to understand further the logic behind decision-making (Boland 2008) in software startups. As a framework, we employ two entrepreneurial logic theories from entrepreneurship literature to investigate how requirement engineering, software design, construction, testing, and software development happen. To the best of our knowledge, this is one of very few attempts to incorporate entrepreneurial logic in the context of software development (Nguyen-Duc et al., 2017; Hevner and Malgonde, 2019). Of previously published studies, we are aware only of Khurum et al.'s (2015) use of the opportunity recognition theory and Hevner and Malgonde's (2019) assessment of effectuation theory in platform development. Unlike Hevner, we describe both effectual and causal logics in each SE activity. We also propose an explanatory model of the influences of entrepreneurial logic on software development activities in startups. This study aims to better understand the connections between the logic of startup founders and SE activities. Two research questions (RQs) were derived from the research objective:

RQ1: How do entrepreneurial logics apply to SE activities in startups?
RQ2: How do entrepreneurial logics apply to software product development at the company level?

The remainder of the paper is structured as follows: Section 2 contains background and related work, Section 3 explains the research method, Section 4 describes the results, Section 5 describes the findings, and Section 6 concludes the paper.

## 2. Related Work

The section presents important definitions used in this paper, background and related work about Software Startups, Software Engineering in Startups and Entrepreneurial logics. The key terminologies are summarized in Table 1.

Table 1 Key terminologies

| Terms | Definitions | Reference |
|---|---|---|
| Software startup | Highly reactive and rapidly evolving software-intensive product development companies with an innovation focus and a lack of resources, working under uncertainty and time pressure | Section 2.1 |
| Startup stage | Three main stages are pre-startup, startup and post-startup | Section 2.1. |
| Lead Users | Users who have a needs of general market but earlier than the crowd | Section 2.2 |
| Minimum Viable Product | A version of a product with just enough features to be usable by early customers | Section 2.2 |
| Entrepreneurial logic | The process of creatively defining, reframing and taking action to make sense out of business situations | Section 2.4 |
| Sense making | A process by which people give meaning to their collective experiences | Section 2.4 |
| Causal Logic | A process of pursuing a predetermined goal by acquiring needed resources, tools to achieve the goal | Section 2.4.1 |
| Effectual Logic | A process of selecting among several possible goals with a pre-given set of resources | Section 2.4.2 |
| Technical debt | Implied cost of additional rework caused by choosing a quick technical solution to meet an urgent demand instead of a sustainable approach that would take longer. | Section 2.5 |

### 2.1. Definitions of Software Startups

The term "startup" has been defined differently across various disciplines (Sutton, 2000; Ries, 2011; Blank, 2013; Unterkalmsteiner et al., 2016; Ghezzi, 2018; Steininger, 2019). Steve Blank (2013) describes a startup as a temporary organization that aims to create innovative high-tech products without a prior working history as a company. The author further highlights that the business and its product should be developed in parallel within the startup context. Eric Ries (2011) defines a startup as a human institution designed to create a unique product or service under extreme uncertainty. Rather than a formal company, a startup should be considered a temporary organizational state that seeks a validated and scalable business model (Unterkalmsteiner et al., 2016). A company with a dozen employees can still be in a startup state while it validates a business model or a market. As previous startup research has done (Berg et al., 2018), we define a startup as a highly reactive and rapidly evolving company with an innovation focus and a lack of resources, working under uncertainty and time pressure. We looked for companies that develop software products as their primary value proposition or include software as a significant part of their products or services.There are many different startup life-cycle models describing startups' states of objectives, resources and business maturities. A startup model can have from three to seven stages, depending on the aspects they focus on. As adopted in our previous work (Nguyen-Duc et al. 2016, 2017), we define startups' phases as the followings:

- Pre-startup stage: ideas are developed and need to be validated, startups in the quest for financial and human resources. Startup activities are carried out by founders or short-term hires. The purpose of this stage is to demonstrate business feasibility, team building and management. The common financing model is bootstrapping, family, friends and foes (FFF)
- Startup stage: prototypes are developed and experimented, startups have already figured out the problem/solution match. Some revenue is generated, but not necessarily over the break-even point. Founder seeks support mechanisms from startup ecosystems, learn to accelerate their business development. The common financing model is own funding and seed funding.
- Post-startup stage: products are extended, startups achieve the product/market match. Startups expand their customer bases, the revenue models are predictable and scalable. A hierarchical structure is formed within the startups. The common funding model is Series A, Series B, and other series

### 2.2. Agile development, User-centered Design and Lean startups

Contributions to agility and reactiveness of product development are known from Agile (Beck et al., 2001), Lean (Gautam and Singh, 2008; Ries 2011), and User-centered Design (Norman 1986; Gothelf 2013) methodologies. Dealing with certain levels of uncertainties can be seen from different agile practices, such as short development cycles, collaborative decision-making, rapid feedback loops, and continuous integration enable software organizations to address change effectively (Highsmith and Cockburn, 2001; Beck and Andres, 2004). In startup contexts, Giardino et al. showed that agile practices are adopted, but in an ad-hoc manner (Giardino et al., 2014). Pantiuchina et al. studied 1256 startup companies and reported that different agile practices are used to different extents, depending on the focus of the practices (Pantiuchina et al. 2017). The authors found that speed-related agile practices are used to a greater extent in comparison to quality-related practices. Recently Cico et al. reported that startups in their growth phases do apply Agile practices in various ways. Strict adoption of agile methodology seems not to be perceived critically, and in some situations, it is difficult to apply agile practices due to the nature of developing products (Cico et al., 2020).

Lean startups with the focus on forming hypotheses about businesses, building experiments to evaluate them (Ries 2011), had a large impact on startup and research communities. Minimum Viable Product (MVP) is a central concept of the approach, defined as a version of a product with just enough features to be usable by early customers (Ries 2011). Bosh et al. discussed why few practitioners apply Lean Startup methods because of the lack of guidelines for method operationalization (Bosch et al., 2013). Other factors influencing the implementation of Lean Startup are also reported, such as the costs of prototyping in particular (Ladd et al., 2015), experience and knowledge about the methodology (Nguyen-Duc et al., 2016, 2017), and experimentation in general (Gemmell et al., 2012).

Customer Development is another popular paradigm that focuses on customers upfront, i.e. developing the customers rather than products in the early stages of startups (Blank 2007; Blank & Dorf, 2012; Alvarez 2014). So startups are advised to search for the right customers to test their business hypotheses and thus obtaining validation or refutation of the overall business model. This relates to the marketing practices of lead users who (1) face the needs that will be general in the market, but face them much earlier than the crowd, (2) are positioned to benefit significantly by obtaining the solution to those needs (von Hippel, 1986). User-Centered Design is also a relevant paradigm for certain types of startups, as they aim for creativity and empathy for designing user-centric solutions and helping developers to change their mindset on how to approach a problem and envision its solution (Signoretti et al., 2019). Hokkanen et al. studied User Experience (UX) practices in startups and suggested that startup products need to fulfill minimal functional and user experience requirements (Hokkanen et al., 2015).

Startups, in general, do not follow one or many of these methodologies strictly. This applies not only to startup companies, as a recent large-scale survey in European software companies showed that modern software and system development does not follow any blueprint and adopt different hybrid approaches (Tell et al, 2017). The understanding of which compositions of development methods, i.e. Agile, Lean, etc that actually work in software development contexts is missing (Tell et al, 2017). In this work, we aim to understand the possible links between adopted development practices with the entrepreneurial logics.

### 2.3. Software Engineering Models for Startups

The need for understanding and modeling SE phenomenon in startup companies has been recognized in SE literature. Giardino et al. (2016) explained a phenomenon of accumulated TD in startup contexts when product quality is a low priority and the startup team is more focused on speeding up development. The authors pointed out that lack of resources is the main driver for the observed product development patterns; however, they did not explain how the limited resource leads to the lack of focus on quality. Nguyen-Duc et al. (2016) described startup development patterns by looking at the co-evolution of product and business as an inter-twined process: startups need entrepreneurial skills and project management skills when hunting implementing opportunities (Nguyen-Duc et al., 2015). These models take into account the influence of business factors in decisions on product development. However, the number of cases investigated at that time was limited. Fagerholm et al. (2017) implemented the Build-Measure-Learn cycles (Ries, 2011) as continuous experimentation systems, where the new product idea can be hypothesized and tested.

Some studies explored particular activities of product development, such as MVP development or requirement engineering. Nguyen-Duc et al. (2017) described how MVPs are used in different software startups, and Tripathi et al. (2018) revealed how the supporting roles of startup ecosystem elements influence MVP development. More recently, Melagati et al. (2019) presented how founders and other factors influence startups' requirement engineering activities. These studies acknowledged the impact of entrepreneurs on SE activities; however, they do not have a theoretical model to explain this impact. These studies call for the adoption of decision-making theories to fill the gap. More recently, Klotins, Unterkalmsteiner, Chatzipetrou, et al. (2019) looked at commonalities among startups' goals, challenges, and practices. The authors showed that startups share the same SE challenges and practices with established companies; however, startups need to evolve multiple activities simultaneously. The study described product development startups from a project management perspective to consider planning, measuring, and controlling activities. This assumption suggests a plan-driven logic when looking at the startup development process and might lead to a similar observation when comparing these plan-based activities to those of established companies. In contrast to a plan-driven and controlled approach, effectual logic adopts means-driven, emergent, and flexible mechanisms to deal with the environment's uncertainty. Previous studies in SE have suggested that the availability of resources can influence the occurrence of engineering phenomena, i.e., TD (Giardino et al., 2016) and the choice of which MVP to implement (Nguyen-Duc, Dahle, et al., 2017).

## 2.4. Effectuation and Causation Logics in Entrepreneurial Decision Making

Entrepreneurial logic is defined as a process of creatively defining, reframing and taking action to make sense out of situations that require new assumptions and understandings (Cunnhingham et al. 2002). Sensemaking is defined as "the ongoing retrospective development of plausible images that rationalize what people are doing" (Weick 1995, Weick et al. 2005). In the purpose of making sense from startup situations, two kinds of entrepreneurial logic that have recently gained research attention are the logics of effectuation and causation (Sarasvathy, 2001; Alvarez and Barney, 2005; Fisher, 2012; Reymen, 2015). Below, we present their definitions and examples in the context of software development.

### *2.4.1. Causal Logic*

In a nutshell, causal logic describes a process of pursuing a predetermined goal by acquiring needed resources, tools to achieve the goal. The causal logic focuses on the predictable aspects of an uncertain future and follows the logic of "to the extent we can predict the future, we can control it" (Sarasvathy, 2001, p. 7). An example of this approach is to conduct a project in a large company. When a project manager is assigned to the project, he perhaps needs to gather his team to apply for extra resources if needed. He needs to be aware of project constraints and perform to achieve the predetermined goals of the project. The project manager's attitude towards unexpected contingency is avoided. He relies on accurate predictions, careful planning, and focusing on predetermined objectives. In the causation model, startups focus on competition and constrain task relationships with customers and suppliers. For instance, the project manager needs to manage the relationships with external stakeholders to limit their possible negative influences (delays in the project schedule, unexpected costs, and other unanticipated problems). The causation model highlights the action to maximize returns by selecting optimal approaches (Sarasvathy and Dew, 2005b). The manager will prioritize analytical calculations and pursue an optimized approach.

### *2.4.2. Effectual Logic*

An effectual logic describes a process of selecting among several possible goals with a pre-given set of resources (Sarasvathy, 2001; Barney, 1991). The effectual logic focuses on the controllable aspects of an unpredictable future and follows the logic of "to the extent we can control the future, we do not need to predict it" (Sarasvathy, 2001). For example, a startup that is a spin-off from a university has technological patents. The startup decides to develop different business models leveraging the application of the patents. The effectuation-driven startup tends to involve as many people as possible in the early stages to generate value for the startup. Instead of focusing on maximized returns, the effectuation-driven startup examines how much one is willing to lose on a startup journey. In our example, the startup team needs to calculate and commit only the resources, time, and effort that they can tolerate wasting.

## 2.5. The Need for Entrepreneurial Logic in Software Development

Traditional software development approaches start with a particular goal and realize it through a linear or iterative development process, which is largely overlap with causal logic. Significant parts of SE research base on a prescriptive assumption that software development projects can be guided by reference frameworks, processes, techniques, and tools. When managing a software project, one could assume a certain level of control based on plan-driven and systematic working manners (Klotins, Unterkalmsteiner, Chatzipetrou, et al., 2019) where project context, such as market, customers, and other ecosystem elements, are somewhat identified as a priori.

Table 2 Software startup phenomenon and their possible connections to entrepreneurial logics

| Phenomenon | Description | Reference | Judgment |
|---|---|---|---|
| Software pivot | A pivot is a strategic change designed to test a fundamental hypothesis about a product, business model, or growth engine. | Model of pivot triggering factors (Bajwa et al., 2017; Bajwa, | A certain type of product pivot would be desirable and plan-driven. But most of the pivots are triggered by external factors and reflect the effectual logic |
| | | 2020; Khanna et al., 2018) | |
| Technical debt (TD) | implied cost of additional rework caused by choosing a quick technical solution to meet an urgent demand instead of a sustainable approach that would take longer. | Greenfield model of software startups (Giardino, 2016; Seaman and Guo, 2011) | Startups accumulate TD, but its nature might be different from large companies in that TD is a way to manage tolerable loss in effectual logic. |
| Minimum Viable Product (MVP) | A version of the product to collect validated learning | MVP-based learning (Nguyen-Duc et al., 2017; Duc and Abrahamsson, 2016) | Startups develop many MVPs but in order to gather necessary learning, they need a more plan-driven approach |
| Customer involvement in product development | Customers involve early and often in requirement, design, and testing activities | Continuous involvement (Nguyen-Duc et al., 2017; Melegati et al., 2019; Yaman et al., 2016) | Effectuation-driven companies encourage the contribution of external stakeholders in co-creating company value |

Software development in startups often needs to deal with multiple-influenced and rapidly changing business and working environments, which makes effectual logic relevant (Giardino et al., 2014; Giardino et al., 2016; Bajwa et al., 2017). In software startups, product development is often essential for the success or failure of the company. They are often limited in resources and work under pressure to prove their products or services to attract funding. Such settings make formal software development paradigms less applicable (Pantiuchina et al., 2017; Kemell et al., 2019). Notably, previous studies also reported that startups' working way is contingent on their environment (Nguyen-Duc et al., 2015; Nguyen-Duc et al., 2016; Kemell et al., 2019). Effectual logic could help explain decisions or activities taken in resource-constraint situations. Brettel et al. showed that effectuation is positively linked to process output and efficiency in highly innovative RandD projects (Brettel et al., 2012). Similarly, these logics could be relevant to SE activities in startups. Several previously studied technical concepts in software startups, such as MVP and TD, can be explored further under entrepreneurial logic (as shown in Table 2).

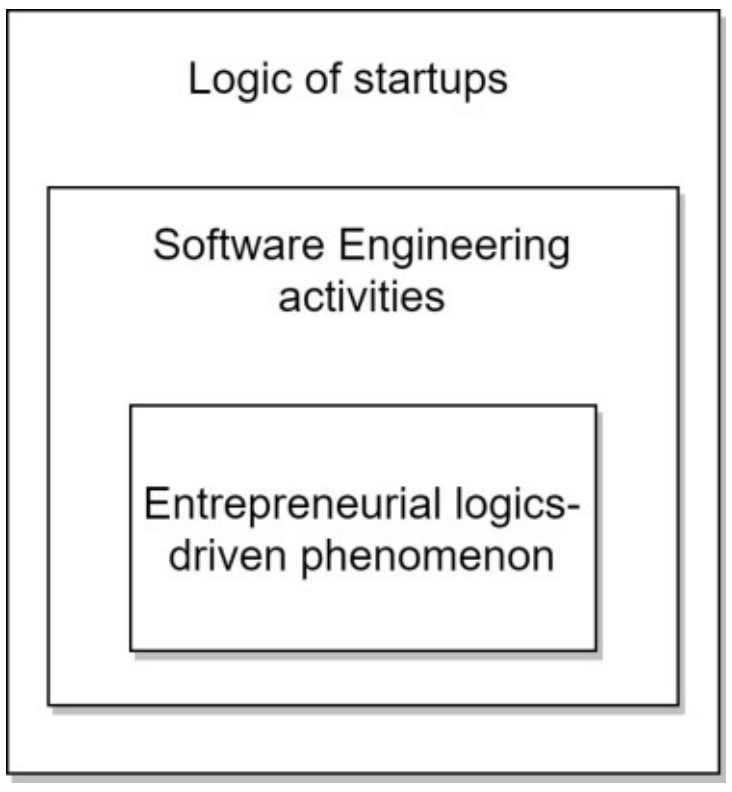

Figure 1 A conceptual framework of entrepreneurial logics in software startups

In this study, we argue that entrepreneurial logic can help to understand specific patterns between entrepreneurial contexts and how product development activities are chosen. Mansoori proposed a three-tiered framework for the mapping of entrepreneurial processes onto the three levels: logic, model, and tactics (Mansoori, 2015, 2020). At the logic level, principles for startups include the notion of uncertainty (epistemological or ontological), view of the future (predictable or completely unknown), nature of the process (discovery or creation), epistemological discussions (realism or constructivism), and relation to external stakeholders (transactional or generative) (Mansoori 2015). At the model level, there are often organized sequences of operations and interactions for guiding entrepreneurial actions. At the tactic level, there are activities, exercises or practices that are in line with the underlying logic and the prescribed model. While the original framework applies to entrepreneurial activities in general, we adopted it to the software engineering model and activities, as seen in Figure 1. At the model level, we relate the entrepreneurial logic to SE processes, i.e. requirement engineering, software design, implementation, and testing. At the tactic level, we will extract specific SE activities and practices that characterize entrepreneurial logic. This study takes the first step towards entrepreneurial software development by exploring the connection between entrepreneurial logic and software product development activities. Our intention is not to predict startups' behaviors or to classify them as either kind of logic.

## 3. Research Methodology

Our research goal was to generate new knowledge about the logic behind software product development in startups, which needs to be investigated in its natural setting. Inductive research with a bottom-up exploration of evidence and conclusions generated from this evidence is a suitable approach commonly adopted in empirical SE research (Seaman, 1999; Wohlin and Aurum, 2015; Ayala et al., 2018; Khurum et al., 2015). All the different research methodologies have their place in software engineering, and each approach has value for the software engineering practitioner (Easterbrook et al., 2008). The possible choices for such an empirical study were exploratory, descriptive, explanatory, and evaluation research (Collis and Hussey, 2009). Compared to other SE research fields, software startup research is still a growing field with a limited understanding of engineering-specific activities in this context (Unterkalmsteiner et al., 2016; Berg et al., 2018; Klotins et al., 2019). From philosophical perspectives, the study adopted a mixed view between interpretivism and positivism. On one hand, the study has many assumptions of interpretivism, i.e. research must be interpreted within the context in which it takes place, and research findings are subjective (Walsham, 1995). Besides, the goal of this research is to provide deep insight regarding entrepreneurial scenarios, not to confirm or test a hypothesis. Therefore, we endeavored to explore software development from entrepreneurial perspectives descriptively; we explain how to plan and organize startups' work accordingly. Since these phenomena involve mainly human factors, it is vital to evaluate human perceptions of the subject (Easterbrook, 2008). On the other hand, we adopt the concept of cognition from positivism, emphasizing the role of empirical evidence in the formation of ideas, rather than innate ideas or traditions. Systematic approaches to collect and analyze evidence is pursued towards reproducible findings and logic-based science. By collecting data in the form of responses to standardized questions, i.e. survey research, accumulated evidence can constitute facts.

It is not uncommon in SE/Information Systems (IS) research for empirical studies adopting both paradigms (Runeson and Höst, 2009, Stol et al., 2016). In a mixed-research approach, qualitative data can be coded quantitatively" by counting words and categorizing statements (Trochim, 2001) or combinations of survey data with case studies (Ralph, 2015). We have adopted the approaches in our previous work (de O. Melo et al., 2013; Ayala, 2018). To gather and interpret the evidence needed to answer our research questions, we conducted semi-structured interviews with startup cases. Depending on the in-depth knowledge of a case, qualitative research can focus narrowly on a few case studies, or tackle a broader scope. We used the same set of key questions repeatedly in a relatively large number of cases (N=40). We aim at observing both frequency distribution and systematic and thematic patterns across interview cases.
This study could have been carried out at the activity, team, project, and company levels. To associate entrepreneurial logic with SE, we needed to look at specific activities and their context; hence, the first analysis was performed at the activity level (RQ1). Since we collected data from different companies, it was straightforward to then perform the second analysis at the company level: in other words, a cross-case analysis.

### 3.1. Case Selection

The challenge of identifying proper startup cases and differentiating the similar phenomena represented among them — freelancers, SMEs, or part-time startups — is well known in software startup research (Unterkalmsteiner, 2016; Berg et al., 2018). Based on the successful approaches adopted in previous studies (Klotins, Unterkalmsteiner, Chatzipetrou, et al., 2019; Berg et al., 2020), we defined five criteria for our case selection:

- A startup that has at least two full-time members, so their MVP development is not individual activities
- A startup that operates for at least six months, so their experience can be relevant
- A startup that has at least a first running prototype, so the prototyping practice is a relevant topic
- A startup that has at least an initial customer set, i.e., first customer payments or a group of users, so that certain milestones in the startup's process are made
- A startup with software as the central part of its business core value

We intended to conduct multiple interviews in each startup to achieve data triangulation (Boyatzis, 1998); however, most startups could only provide a single interview. We obtained multiple follow-up interviews in seven cases (S01–S05, S07, S08), which provide the main insights. The other 33 cases, with a single interview, extend, and confirm the findings from the principal cases. All the information about the cases was collected via internet research and written documents provided by the companies to address gaps left by the lack of follow-up interviews. The characteristics of the cases studied are summarized in Section 3.3.

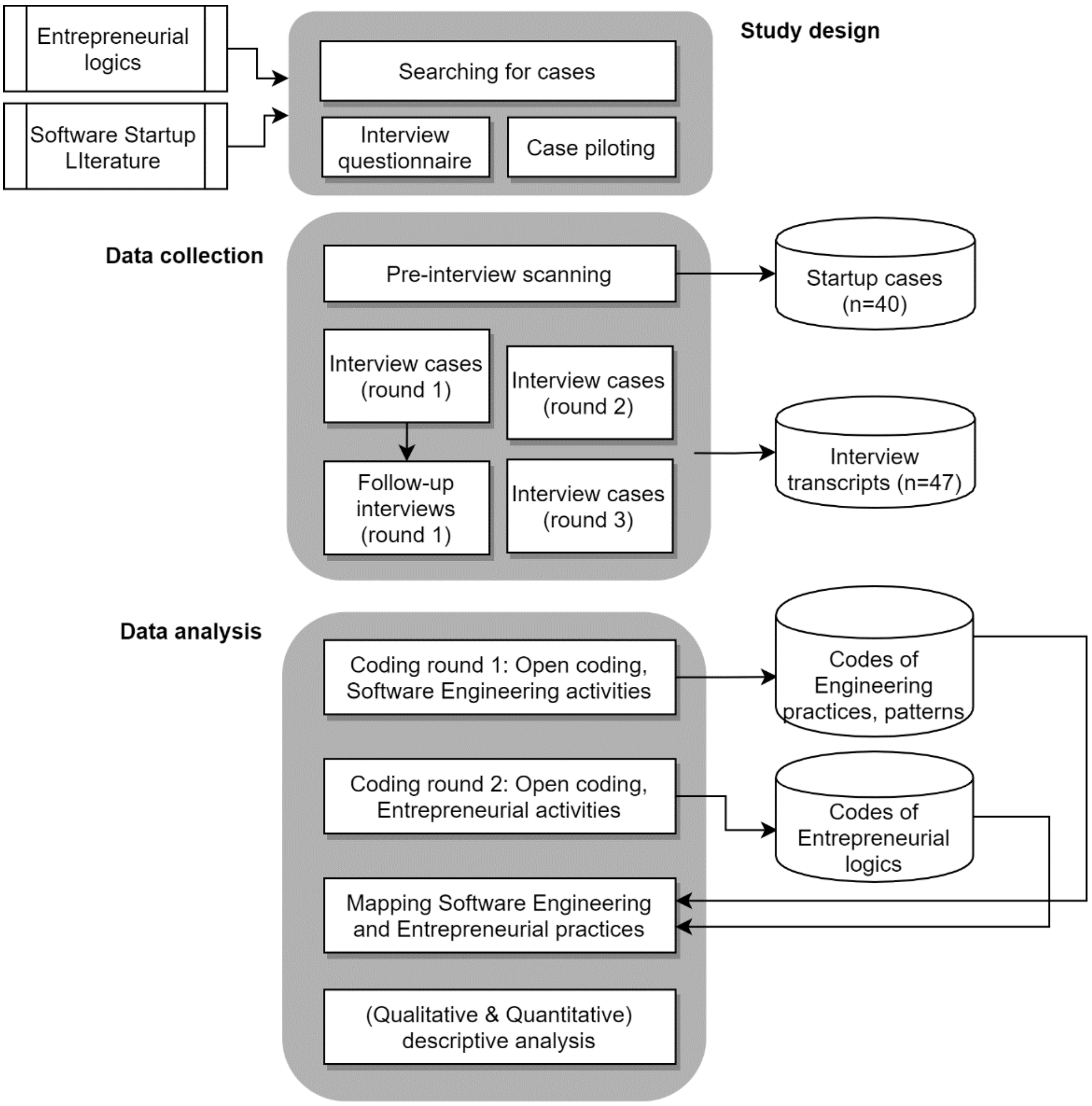

Figure 2 Data collection and analysis process

### 3.2. Data Collection

The main data collection methods were semi-structured interviews, participant observations, face-to-face discussions with project leaders, and document analysis. The identification and collection of data were performed in three rounds. The first round was conducted from March 2015 to February 2016; data collection was mainly done by the first author and a research associate. The second round was conducted during September 2016 and January 2017. The third round was conducted from September 2017 to June 2018; this data collection was performed collectively by the first author and graduate students at the Norwegian University of Science and Technology. A consistent approach was undertaken to collect data (as shown in Figure 2). The data collection process was as follows:

Step 1: Identifying cases. Contacts for startups were searched via four channels: (1) startups within the authors' professional networks; (2) startups in the same towns as the authors and from Startup Norway; and (3) startups listed in the Crunchbase database. We also included contacts we made at startup events, such as the Norwegian Investment Forum, Startup Weekend, and Hackathons.

Step 2: Feasible analysis. We spoke with software startups in coworking spaces and incubators in Trondheim, Norway, to become familiar with startup scenes and their current issues.

Step 3: Study design. Several more interviews were conducted both face-to-face and remotely to build data collection equipment. The interview guideline was modified from an existing one, which focused on the topic of startup pivots.

Step 4: Case piloting. Case analysis was conducted using information available from the internet or provided by the case companies that allowed a holistic understanding of each case and provided more substantial evidence for the conclusions drawn from the interviews. This step was conducted before proceeding to the actual interview with startups.

Steps 5 and 6: Data collection. Interviews allowed us to collect information in the participants' own words rather than by limiting them to predefined response choices on a survey (Oates, 2005). We chose to conduct semi-structured interviews, as these are expected to give a researcher the flexibility to probe deeper into unforeseen information that may emerge during interviews (Seaman, 1999). Each interview lasted between 40 and 70 minutes. The number of interviews in each round is shown in Table 3.

Table 3 Data collection rounds

| **Round** | **No. of contacts** | **No. of cases** | **No. of interviews** |
|---|---|---|---|
| 1 | 219 | 20 | 27 |
| 2 | 40 | 7 | 7 |
| 3 | 47 | 13 | 13 |

The final contact list included 306 startups from the USA, Norway, Sweden, Finland, Italy, Germany, Spain, the Netherlands, Singapore, India, China, and Vietnam. We approached the companies on the final list to search for participants; many startups expressed their interest in the study results but did not have time to participate. These companies responded to our call for participation with sentiments similar to "[t]he research appears interesting and relevant to our experience. Unfortunately, we do not have the resources and time to participate in such a survey." Besides some large companies who were not interested in the research, we did not see the difference between the ones who accepted and the ones who refused to participate. By emailing and talking via phone, professional networks and nearby startups have a slightly lower turn over rate than startups from Crunchbase. However when approaching startups via personal introduction or meet in person, there is significantly higher chance to aquire their participation.

Excluding startups that were not interested in the research or startups that did not meet our selection criteria, the final number of eligible cases was 40 startups (turnover rate ca. 13%). Among them 25% of the total number of cases come from our convenient networks, 70% of the cases are systematically selected and collected from physical interviews, 5% of the cases are from the CrunchBase database. Some startups required that the authors sign a non-disclosure agreement with the companies; this step was essential to establish a formal link between the researchers and the participating startups and ensure the data confidentiality the companies required to feel more comfortable with our observations of their internal activities.

Table 4: (Common parts of ) the interview guidelines

| |
|---|
| Section 1: Business background<br>Please tell us about your product and your company<br>How was the current software product developed ?<br>What is your team competence? How is it evolved over time?<br>What is your current market?<br>What is your business model?<br>Section 2: Idea visualization and prototyping<br>Could you tell us about the time when: (1) the first idea came to your mind, (2) the first prototype completed, (3) the first payment customer<br>How can you achieve the problem/solution fit with your prototype?<br>Section 3: Product development<br>How many times have you changed? About the most significant pivot: How decisions are made?How was the current software product designed? What is the most challenging issue?<br>How was the current software product implemented and tested? What is the most challenging issue?<br>How was the current software product maintained and extended? What is the most challenging issue? |

In the first round of data collection, most participants answered a simple pre-interview questionnaire in which they filled out basic information about themselves and the company. In some cases, accessing sprint planning documents, product specifications, pitching slides, and communication mailing lists extended our knowledge about startup product development activities. Participant observation occurred in cases S02 and S03, where the first author involved in these cases was either a consultant or a co-founder. These measures facilitated more efficient interviews, as the first author possessed more knowledge about the case and could use less time on formalities. Most of the interviews were conducted by the first author. The author also took notes to mark essential concepts that came up in the interviews. Later on, all the interviews were transcribed using a freelancing service. A researcher in our network recommended the service, and the pilot test of the service was conducted before the study adopted it. The total number of transcripts was 313 A4 pages. In the second and third rounds of data collection, the first authors attended some interviews. Most of the interviews in these rounds were conducted by either graduate students or associated researchers. Although interview questions were slightly different among the three rounds of interviews, the interview structure and key questions remained the same. The interviewees were typically asked about (1) the business background, (2) idea visualization and prototyping, and (3) product development. The common key questions is described in Table 4.

### 3.3. Case Demographics

As shown in Figure 3, our cases vary significantly in terms of application domains. The investigated companies deliver software platforms in healthcare, information technology infrastructure, education, logistics, sales, and marketing. Investigated MVPs included software-intensive products (e.g., mobile apps, dynamic webs, and data analytics) and hardware-relevant products (e.g., Internet of Things platforms). The startup cases present a large spectrum of market segments: prominent startups (65%) targeted a niche market, such as hyper-local news readers, a population of high school pupils and college students, IoT product developers, software developers, and sale-intensive organizations; 35% of the cases currently follow a business-to-business (B2B) model; and the rest operate a business-to-customer (B2C) model. From a geographical perspective, the case sample is biased toward startups serving the Nordic and UK markets: these constitute 75% of the study's total cases. Other geographical markets include the USA, Germany, France, the Netherlands, Poland, Singapore, Hong Kong, and Vietnam. In terms of their headquarters' locations, the demographic representation of the case is shown in Figure 4.

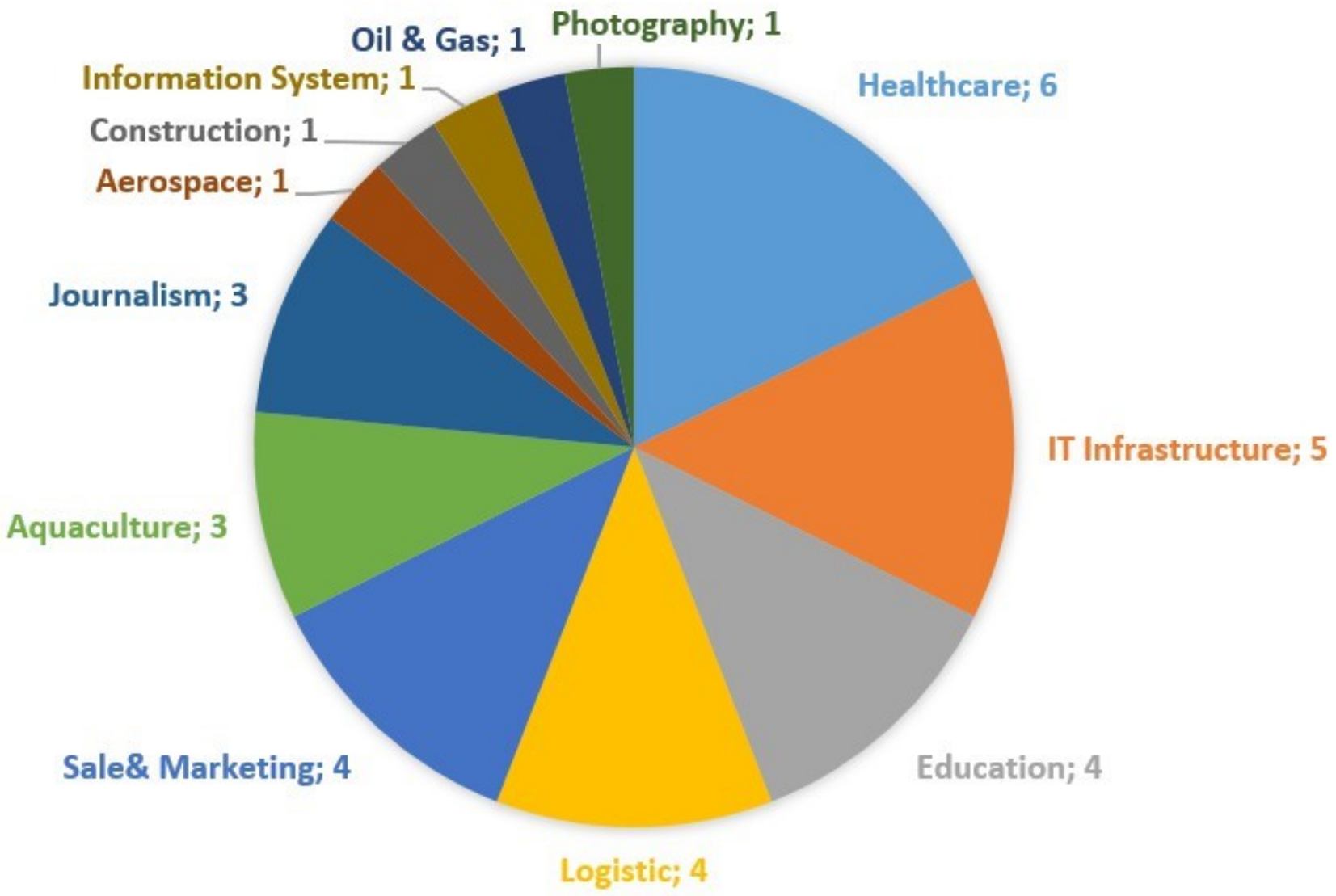

Figure 3 Distribution of startups in application domains

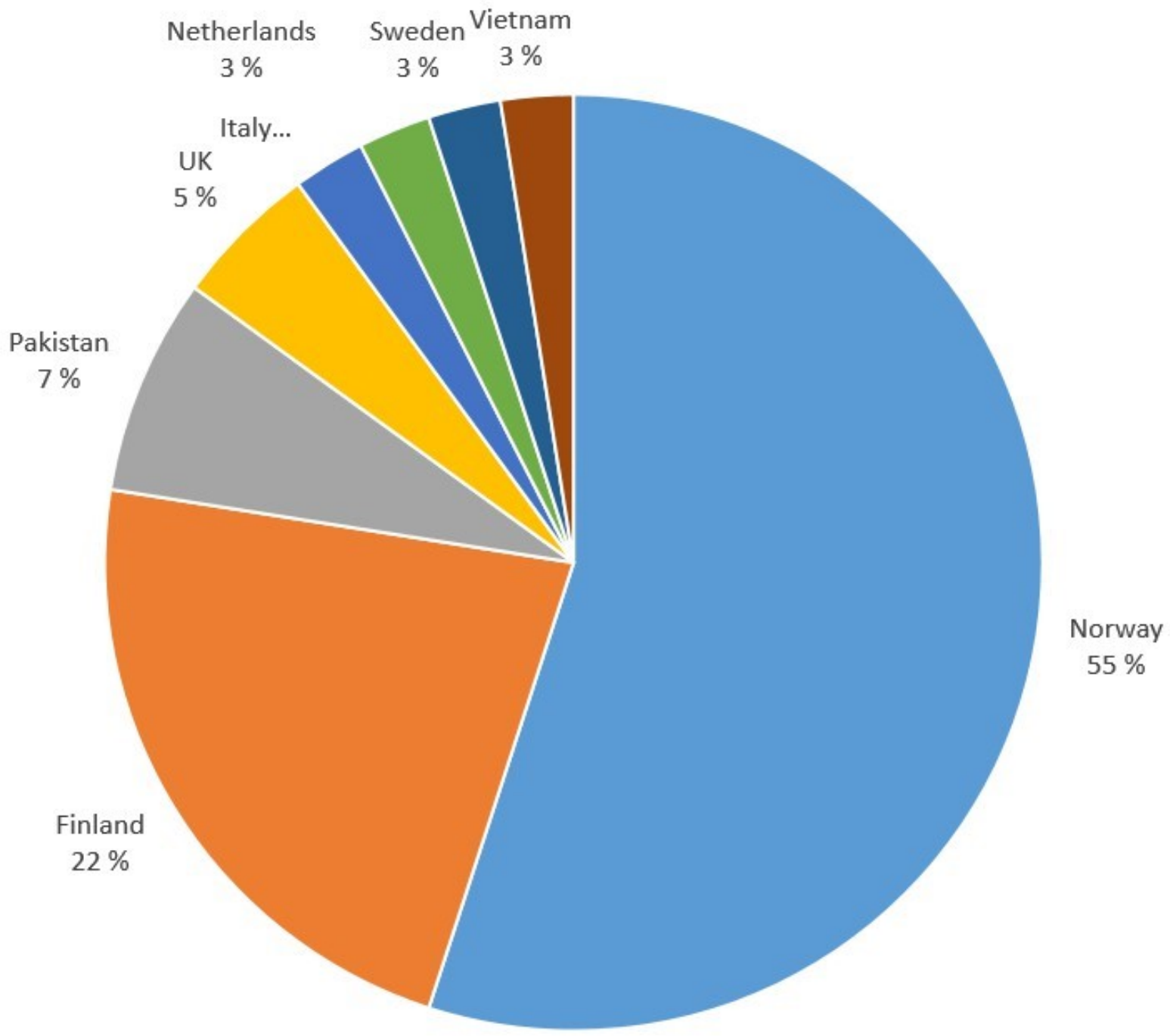

Figure 4 Distribution of startups in terms of headquarters' locations

Team sizes varied from 2 to 85 people, but most of the study's startups (27 out of 40) had a team of between 3 and 20 people. These headcounts include full-time workers employed during the study period, regardless of whether they were included on the payroll at that time. For example, one startup consists of the CEO, CTO, a designer in Norway, and an outsourcing team of six full-time developers in India. The portion of engineers in startups in our sample ranges from 33% to 100%. In many startups, the team is collocated and unstructured. In other cases, it is typicall to observe organizational structures with a separation between product development teams and sales teams. Most of our startups (85% of the total number of cases) are financial bootstraps: they fund the development of products and services through internal cash flow and are cautious with their expenses. Most of the bootstrap startups studied received financial support from their governments, incubators, and accelerators, and startup programs. Some cases were initiated by an investor who secured a stable income for the team. Some cases were in the post-startup phase, with annual revenue above EUR 1 million. Some other startups had invested more than EUR 1 million. The financial information for the remaining startups is either unknown or indicates that they were struggling with their cash flow at the time of the study. The detail profiles of our cases as well as their financial situations are reported in Table 5.

Table 5: The profiles of the startups included in the study

| **Cases** | **Country** | **Product** | **Application domain** | **Years Operational*** | **Current Stage** | **No. People** | **Annual Revenue** | **Source** | **Startup type** |
|---|---|---|---|---|---|---|---|---|---|
| S01 | Italy | Photo trading platform | Arts, Entertainment and Recreation | 4 | Pre-Startup | 15 | <50k Eur | Source1 | 1 |
| S02 | Norway | Hyper-local news platform | Professional, Scientific, and Technical Services | 1 | Pre-Startup | 2 | 110k Eur | Source1 | 2 |
| S03 | Norway | Shared shipping platform | Transportation and Warehousing | 3 | Startup | 6 | <10k Eur | Source2 | 1 |
| S04 | Norway | Digitalized construction management process | Construction | 5 | Post-Startup | 9 | >300k Eur | Source2 | 1 |
| S05 | Finland | Underwater camera product | Agriculture, Forestry, Fishing and Hunting | 4 | Pre-startup | 3 | Unknown | Source2 | 1 |
| S06 | Norway | Sales visualization | Professional, Scientific, and Technical Services | 3 | Post-startup | 18 | >1.5 mil. Eur | Source2 | 2 |
| S07 | Vietnam | Shop location app | Professional, Scientific, and Technical Services | 4 | Startup | 14 | ~200k Eur | Source1 | 1 |
| S08 | Norway | Event and ticket platform | Professional, Scientific, and Technical Services | 4 | Startup | 4 | Unknown | Source2 | 1 |
| S09 | UK | Game-based classroom learning tool | Education | 9 | Post-startup | 12 | ~2 mil. Eur | Source1 | 2 |
| S10 | Norway | IoT OS platform | Professional, Scientific, and Technical Services | 4 | Startup | 3 | >150k Eur | Source2 | 1 |
| S11 | Norway | Ticketing system | Professional, Scientific, and Technical Services | 4 | Startup | 5 | >150k Eur | Source2 | 1 |
| S12 | Norway | eLearning | Education | 8 | Startup | 3 | Unknown | Source2 | 1 |
| S13 | UK | Shipping services | Professional, Scientific, and Technical Services | 2 | Startup | 3 | Unknown | Source3 | 2 |
| S14 | Sweden | Journalism publishing | Professional, Scientific, and Technical Services | 11 | Startup | 16 | Unknown | Source3 | 2 |
| S15 | Norway | Secondhand marketplace | Professional, Scientific, and Technical Services | 6 | Pre-startup | 2 | Unknown | Source2 | 1 |
| S16 | Norway | Smart grid | Utilities | 5 | Startup | 30 | Unknown | Source2 | 1 |
| S17 | Norway | Simulation-based training | Education | 7 | Startup | 7 | Unknown | Source2 | 2 |
| S18 | Holland | Software development services | Professional, Scientific, and Technical Services | 7 | Startup | 5 | Unknown | Source3 | 1 |
| S19 | Norway | Mobile alert services | Professional, Scientific, and Technical Services | 9 | Startup | 5 | Unknown | Source1 | 1 |
| S20 | Norway | eLearning | Education | 14 | Startup | 13 | 350k Eur | Source2 | 2 |
| S21 | Norway | Fish farm tracking system | Agriculture, Forestry, Fishing and Hunting | 1 | Pre-startup | 6 | Unknown | Source2 | 1 |
| S22 | Norway | Networks of connected camera | Professional, Scientific, and Technical Services | 1 | Startup | 10 | Unknown | Source2 | 1 |
| S23 | Finland | Underwater drone | Agriculture, Forestry, Fishing and Hunting | 4 | Pre-startup | 4 | Unknown | Source2 | 1 |
| S24 | Finland | Tracking devices for shipment | Professional, Scientific, and Technical Services | 2 | Post-startup | 85 | >8 mil. Eur | Source2 | 2 |
| S25 | Finland | Muscle operation measure | Health Care and Social Assistance | 2 | Pre-Startup | 20 | 100k+ Eur | Source2 | 2 |
| S26 | Pakistan | Smart home solution | Manufacturing | 2 | Pre-startup | 8 | Unknown | Source1 | 1 |
| S27 | Pakistan | Smart wheelchair | Health Care and Social Assistance | 1 | Pre-Startup | 3 | Unknown | Source1 | 1 |
| S28 | Finland | Connecting healthcare services to home | Health Care and Social Assistance | 5 | Startup | 5 | Unknown | Source1 | 1 |
| S29 | Pakistan | Smart home devices | Professional, Scientific, and Technical Services | 1 | Pre-Startup | 5 | Unknown | Source1 | 1 |
| S30 | Finland | UI framework for mobiles | Professional, Scientific, and Technical Services | 4 | Pre-Startup | 3 | Unknown | Source1 | 1 |
| S31 | Finland | Aeronautical engineering services | Manufacturing | 5 | Startup | Unknown | Unknown | Source2 | 1 |
| S32 | Norway | IoT solution for gas supplier | Utilities | 2 | Startup | 8 | Unknown | Source2 | 1 |
| S33 | Norway | Personal hydration monitoring device | Health Care and Social Assistance | 2 | Startup | 10 | Unknown | Source2 | 2 |
| S34 | Finland | Enterprise information management solution | Professional, Scientific, and Technical Services | 5 | Startup | 10 | Unknown | Source2 | 1 |
| S35 | Norway | Ear device | Health Care and Social Assistance | 2 | Startup | 5 | Unknown | Source2 | 1 |
| S36 | Finland | Wireless earplug with active noise cancelling | Health Care and Social Assistance | 2 | Pre-startup | 10 | 2000 orders/year | Source2 | 2 |
| S37 | Norway | Autonomous drones | Transportation and Warehousing | 3 | Pre-startup | 7 | Unknown | Source2 | 2 |
| S38 | Norway | Sensor-based detecting systems | Health Care and Social Assistancee | 2 | Startup | 7 | Unknown | Source2 | 1 |
| S39 | Norway | Security for IoT | Professional, Scientific, and Technical Services | 2 | Startup | 3 | Unknown | Source2 | 1 |
| S40 | Norway | Drone control glove | Professional, Scientific, and Technical Services | 1 | Pre-startup | 13 | Unknown | Source2 | 2 |

Notation: Source1: startups within the authors' professional networks; Source2: startups in the same towns as the authors and from Startup Norway; Source3: startups listed in the Crunchbase database

### 3.4. Data Analysis

Data analysis included three steps: (1) labeling SE activities, (2) identifying entrepreneurial logics that occurred in each case, and (3) mapping the entrepreneurial logics and SE activities.

#### *3.4.1. Labelling SE Activities*

We applied a thematic analysis, which is commonly seen in empirical SE research (Cruzes and Dyba, 2011). The objective of our thematic synthesis process was to answer the research questions and come up with a model of higher-order themes describing a way of software development in startups. Braun et al. (2006) suggest six steps for a thematic analysis: (1) familiarizing with data, (2) generating initial codes, (3) searching for themes, (4) reviewing themes, (5) defining and naming themes, and (6) producing the report. As suggested in the literature, we adopted open coding (Braun et al., 2006; Wohlin and Aurum, 2015). Sentences that mentioned SE activities and their contexts are labeled. We developed a taxonomy of startup knowledge and practice areas, including SE knowledge areas from SWEBOK (Bourque and Fairley, 2014). We tried to produce as many codes as possible to avoid missing any relevant or interesting information. The coding scheme for SE activities includes:

P0. SE (general)
P1. Requirement Engineering
P2. Product Design
P3. Software Construction
P4. Software Testing
P5. Software Maintenance
P6. Software Process Management

#### *3.4.2. Identifying Effectuation-driven and Causation-driven Behaviors*

To understand the logic behind decisions in a startup, we need first to understand the startup journey and important milestones. The first step was to read through the transcribed interviews to generate initial ideas and identify possible trends or patterns. For each case, we extracted texts related to critical events that occurred during the startups' journeys. Our approach is similar to previous studies that used key events identified through information from the interviews (de O. Melo et al., 2013; Reymen et al., 2015; Reymen et al., 2017; Fagerholm et al. 2017). Key events were defined as actions or decisions taken by the entrepreneurial teams to create the venture (Reymen et al., 2017). Examples of these events were introducing the first product idea, acquiring funding, initiating collaboration with a supplier, developing the first MVP, product demonstration and launch, hiring employees, and significant pivoting (Reymen et al., 2015). Such events were collected from critical people (e.g., CEOs or CTOs in the startups) and reflected their intentions. The decisions made by external stakeholders were placed in the context category. For each case, we tried to capture each event's timestamp—the pre-startup, startup, and post-startup phases—to plot each case's story in chronological order.

We attended to describe the startup event from participants' perspectives or views. We look for the meaning behind startups' events and activities. Daher stated that "the study of meaning does not directly refer to actual experience, but to the way the self considers its past experience" (Daher et al., 2017). And the meaning of being effectuation-driven or causation-driven is reflected from what the interviewees' own wills.
Identifying entrepreneurial logic was crucial to categorizing a case as either effectuation or causation. As in previous work, we coded entrepreneurial logic at the company level: we created a balanced coding scheme consisting of two theoretical categories based on effectuation and causation theory, i.e., one effectuation and one causation category with four dimensions for each category (Chandler et al., 2011; Reymen et al., 2015). We reused a set of empirical indicators (Sarasvathy, 2001; Read et al., 2009; McKelvie et al., 2020) and modified the coding scheme to make it relevant to the SE context. The coding scheme for the effectual logic includes:

**E1 Basis for acting: means-oriented**
E1.1 Building product mainly on an internal knowledge base and external existing owned resources
E1.2 Defining a general product development plan without concrete details
E1.3 Using internal or resource and infrastructure in the local environment
E1.4 Decisions mainly based on personal preferences
E1.5 Opportunities, ideas, and requirements come from existing contacts
**E2 Attitude towards unexpected events: leverage**
E2.1 Accepting and incorporating unexpected changes, ready for pivots
E2.2 Changing and adapting any potential plans made to accommodate unforeseen events
E2.3 Actively exposing to external stakeholders with an open mind

E2.4 Positively reacting to and incorporating unforeseen developments

**E3 Attitude towards outsiders: partnerships**

E3.1 Reaching trust-based flexible stakeholder agreements and commitments
E3.2 Co-create business with stakeholders
E3.3 Engaging in stakeholder collaborations to pursue opportunities
E3.4 Exposing MVPs to potential clients early on

**E4 View of risk and resources: affordable loss**

E4.1 Be willing to make affordable personal sacrifices (including nonmonetary)
E4.2 Finding unused resources in a local environment (including subsidies)
E4.3 Investing limited, small amounts of personal money, time, and effort
E4.4 Managing growth expectations and ambitions
E4.5 Limiting stakeholders' commitments to levels that are uncritical to them

The coding scheme for the causal logic includes:

**C1 Basis for acting: goal-oriented**

C1.1 Base actions upon expectations (market, technology, policy trends) and predictions (of founders, board members, investors)
C1.2 Defining and pursuing project goals, product, customer needs, or market goals (more specific than "profit" or "a better planet")
C1.3 Defining and satisfying organizational needs (personnel, organization structure, infrastructure, or technology) and selecting between options based on specific goals
C1.4 Evaluating planned progress and adapting means based upon feedback
C1.5 Searching and selecting contacts, clients, and partners based upon predefined plans

**C2 Attitude towards unexpected events: avoid**

C2.1 Feeling threatened by unexpected events, therefore working in isolation with external environments as much as possible)
C2.2 Carrying out plans as defined in cases of unforeseen developments (avoid changes)
C2.3 In cases of unforeseen changes, focusing on activities within startups rather than engaging in environmental factors
C2.4 Pulling away from the project or resolving it quickly in cases of unforeseen developments

**C3 Attitude towards outsiders: competitive analysis**

C3.1 Acquiring resources through market transactions or contract-based agreements with stakeholders
C3.2 Creating and carrying out the patent strategy
C3.3 Carrying out competitor analysis and competitive positioning

**C4 View of risk and resources: expected returns**

C4.1 Maximizing personal profit
C4.2 Calculating and evaluating expected outcomes and returns
C4.3 Planning development in big steps and with large sums (including large recruitment, where *large* is relative for each company)
C4.4 Postponing stakeholder (including clients) contact at the expense of own funds (focus on internal development)
C4.5 Search for stakeholders that commit the amounts necessary for the execution of the plan

To generate initial codes, the first and second authors applied a descriptive coding technique to identify entrepreneurial logic dimensions systematically across all cases (Runeson and Höst, 2009). Descriptive coding helped organize and group similar data into categories, which was the first step towards creating themes. All events in each case were coded according to four effectual and four causal dimensions; thus, effectual, and causal logic could co-occur in the same event. The number of quotes per code is presented in Table 6. We counted how many effectuation dimensions (potentially ranging from 0 to 4) and how many causation dimensions (potentially ranging from 0 to 4) were coded per event. If at least one effectuation or causation dimension was coded for each event, we could identify that event's entrepreneurial logic.

Table 6: Number of quotes per code

| Themes | | No. of quotes |
|---|---|---|
| C1.1 | Base actions upon expectations and predictions | 25 |
| C1.2 | Defining and pursuing project goals | 34 |
| C2.1 | Carefully interacting with environment for secrecy reasons (feel threatened by unexpected events, therefore work in isolation as much as possible) | 12 |
| C2.2 | Carrying out plans as defined in cases of unforeseen developments | 15 |
| C2.3 | In cases of unforeseen changes, focusing on activities within startups rather than engaging in environmental factors. | 11 |
| C2.4 | Drawing back from project or quickly resolving in cases of unforeseen developments | 16 |
| C3 | Attitude towards outsiders. competitive analysis | 19 |
| C4.2 | Calculating and evaluating expected outcomes | 9 |
| C4 | View of risk and resources: expected returns | 5 |
| C4.3 | Planning development in big steps and with large sums | 26 |
| C4.5 | the execution of the plan | 9 |
| E1 | Basis for acting Means-oriented | 49 |
| E1.1 | Building on own knowledge base and other available existing own resources | 15 |
| E1.2 | Short-term planning | 61 |
| E1.3 | Local infrastructure and inside environment | 12 |
| E1.4 | Following personal preferences | 71 |
| E2.1 | Accepting, gathering and incorporating unexpected, leading to pivots | 56 |
| E2.2 | Changing and adapting any potential plans made to accommodate unforeseen events | 26 |
| E2.3 | Actively exposing to outside influences, while being open minded | 33 |
| E3.1 | Reaching trust-based flexible stakeholder agreements and commitments | 20 |
| E3.2 | Co-create business with stakeholders | 24 |
| E3.4 | Exposing MVPs to potential clients early on | 27 |
| E4 | View of risk and resources: affordable loss | 17 |
| E4.1 | Be willing to make affordable personal sacrifice (including non- monetary) | 9 |
| E4.2 | Finding unused resources in local environment (including subsidies) | 9 |
| E4.3 | Investing limited, small amounts of personal money, time and effort | 21 |

We then counted the number of events belonging to either effectuation or causation for each case. In total, we coded 631 events from all cases. The number of coded events varied significantly among cases, from 4 to 26 events in a single case. This variance is due to the relevancy of cases and events in each case to entrepreneurial logic. The total number of effectuation codes is 450 (71.4%). The total number of causation codes is 181 (28.6%).

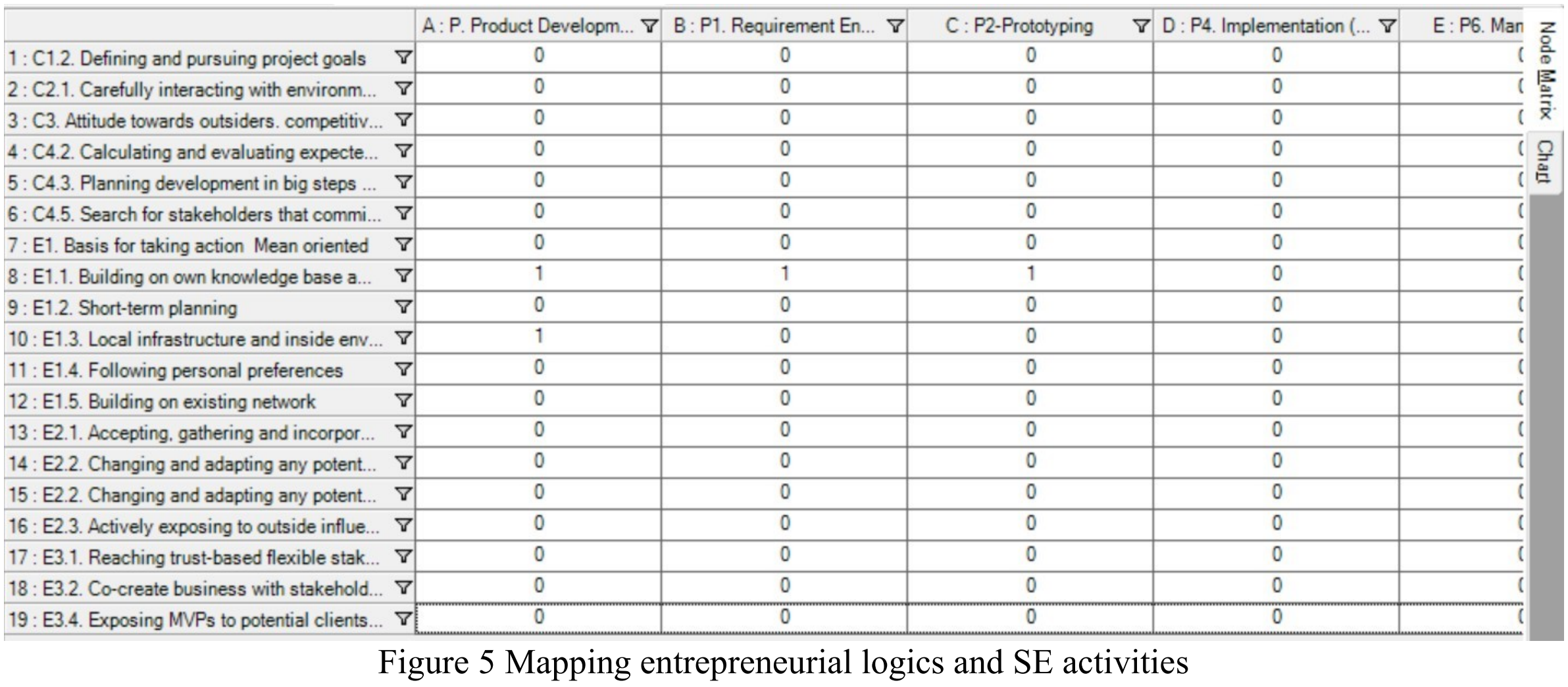

| | A : P. Product Developm... | B : P1. Requirement En... | C : P2-Prototyping | D : P4. Implementation (... | E : P6. Man |
|---|---|---|---|---|---|
| 1 : C1.2. Defining and pursuing project goals | 0 | 0 | 0 | 0 | ( |
| 2 : C2.1. Carefully interacting with environm... | 0 | 0 | 0 | 0 | ( |
| 3 : C3. Attitude towards outsiders. competitiv... | 0 | 0 | 0 | 0 | ( |
| 4 : C4.2. Calculating and evaluating expecte... | 0 | 0 | 0 | 0 | ( |
| 5 : C4.3. Planning development in big steps ... | 0 | 0 | 0 | 0 | ( |
| 6 : C4.5. Search for stakeholders that commi... | 0 | 0 | 0 | 0 | ( |
| 7 : E1. Basis for taking action Mean oriented | 0 | 0 | 0 | 0 | ( |
| 8 : E1.1. Building on own knowledge base a... | 1 | 1 | 1 | 0 | ( |
| 9 : E1.2. Short-term planning | 0 | 0 | 0 | 0 | ( |
| 10 : E1.3. Local infrastructure and inside env... | 1 | 0 | 0 | 0 | ( |
| 11 : E1.4. Following personal preferences | 0 | 0 | 0 | 0 | ( |
| 12 : E1.5. Building on existing network | 0 | 0 | 0 | 0 | ( |
| 13 : E2.1. Accepting, gathering and incorpor... | 0 | 0 | 0 | 0 | ( |
| 14 : E2.2. Changing and adapting any potent... | 0 | 0 | 0 | 0 | ( |
| 15 : E2.2. Changing and adapting any potent... | 0 | 0 | 0 | 0 | ( |
| 16 : E2.3. Actively exposing to outside influe... | 0 | 0 | 0 | 0 | ( |
| 17 : E3.1. Reaching trust-based flexible stak... | 0 | 0 | 0 | 0 | ( |
| 18 : E3.2. Co-create business with stakehold... | 0 | 0 | 0 | 0 | ( |
| 19 : E3.4. Exposing MVPs to potential clients... | 0 | 0 | 0 | 0 | ( |

Figure 5 Mapping entrepreneurial logics and SE activities

### *3.4.3. Mapping Entrepreneurial Logics and SE Activities*

Not every event relates to SE. We went through each case and identified the quotes that had both SE labels and entrepreneurial logic labels. An example of how two layers of codes are matched is shown in Table 7, with some quotes extracted from case S12. To understand how quotes about SE activities are related to entrepreneurial logic, we employed axial coding (Corbin and Strauss, 1990) to map them into entrepreneurial logic coding scheme.

Table 7 Startup behavior quotes from case S12

| SE Area | Quotes | Entrepreneurial Logic |
|---|---|---|
| P1. Requirement Engineering | Either we solve them by providing them different products or we do ignore parts of the market. We make a very active statement on what kind of requirements we do fulfill. Then we turn down clients that do not believe the [00:17:48] requirement. We make a very clear statement to what we think the future of journalism is, then we pursue it. The cost of that is neglecting parts of our market. | C4.5. Search for stakeholders that commit the amounts necessary for the execution of the plan |
| P1. Requirement Engineering | That is because we are in a very challenging market with changing requirements, so that is what they want. Then, as we got bigger, we tried to create a more complex organization within the company. That was the biggest challenge, or at least to us, because we did not know how to do it. | E1. Basis for acting: means-oriented |
| P1. Requirement Engineering | There will always be requirements arriving, that is one thing. Sometimes the new requirements disrupt the old requirements. At the moment, we are working to disrupt the old products. To reinvent them and to kick the [00:15:36] away under our old products. | E2.2. Changing and adapting any potential plans made to accommodate unforeseen events |
| P6. Process Management | Yes, we have always been working in an agile Way. We are not adhering to any specific agile approaches, but we can't do long-term specifications. That is not doable in an industry that is changing very rapidly. We have always been working with long-term visions but with short-term specifications. The way we developed specifications, it is always with the collaboration with the clients or the customers | E3.4. Exposing MVPs to potential clients early on |
| P3. Software Construction | We do all software development in-house, we do not do any outsourcing to India or other places for the simple reasons that everything we do is very short cycle. It's very innovation-oriented, so our software developers probably taught 50% of the time and code 50% of the time, so outsourcing wouldn't really work for the way we work | E1.1. Building on their knowledge base and other available existing owned resources |

Two-dimensional queries were created in NVivo version 12 to map the entrepreneurial logics and SE activities, as shown in Figure 5.[4] To aid the mapping process, we developed a qualitative codebook that includes all quotes and their associated codes (illustrated by Figure 6). Two authors read the codes, the case context and assign an explanation to them. We use collaborative notes and mind maps as additional tools to record any discoveries in the data.

### *3.4.4. Effectuation Index for Cases*

To determine whether a case is either effectuation or causation dominant, we defined an Effectuation Index (EI), which has been used in a previous study (McKelvie et al., 2020):

EI = X(EffectuationEvents)/ X(Events) (1)

For comparison among cases, we defined three categories based on the value of EI:

- EI between 0.7 and 1: effectuation dominant

[4] https://www.qsrinternational.com/nvivo-qualitative-data-analysis-software/home

- EI between 0.3 and 0: causation dominant
- EI between 0.31 and 0.69: mixed

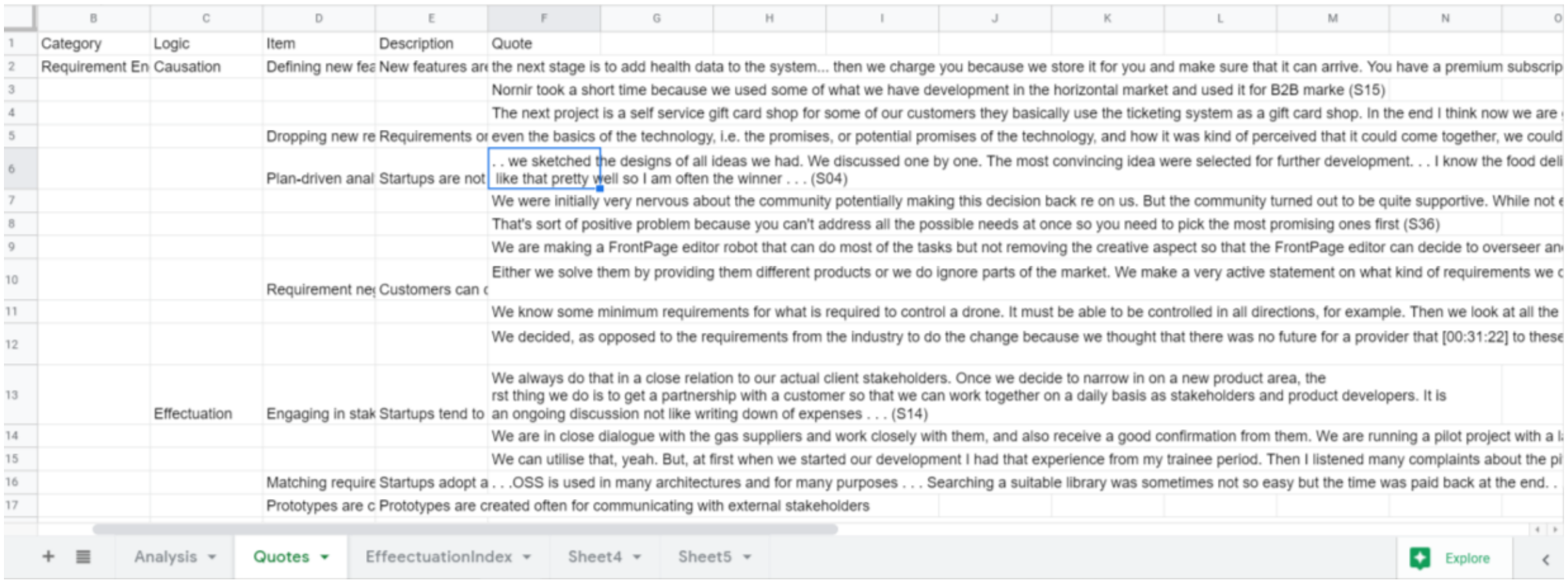

Figure 6 An example of the qualitative codebook

Table 8: Effectuation Index value of each case

| Id | No. Events | No. Effectuation. | No. Causes. | Effectuation Index | Logics type |
|---|---|---|---|---|---|
| S01 | 8 | 7 | 1 | 0,87 | Effectuation |
| S02 | 13 | 9 | 4 | 0,69 | Mixed |
| S03 | 21 | 15 | 6 | 0,71 | Effectuation |
| S04 | 19 | 14 | 5 | 0,74 | Effectuation |
| S05 | 7 | 6 | 1 | 0,86 | Effectuation |
| S06 | 12 | 5 | 7 | 0,42 | Mixed |
| S07 | 7 | 6 | 1 | 0,86 | Effectuation |
| S08 | 8 | 7 | 1 | 0,87 | Effectuation |
| S09 | 14 | 6 | 8 | 0,43 | Mixed |
| S10 | 8 | 7 | 1 | 0,87 | Effectuation |
| S11 | 19 | 14 | 5 | 0,74 | Effectuation |
| S12 | 6 | 5 | 1 | 0,83 | Effectuation |
| S13 | 15 | 9 | 6 | 0,6 | Mixed |
| S14 | 16 | 11 | 5 | 0,69 | Mixed |
| S15 | 16 | 14 | 2 | 0,87 | Effectuation |
| S16 | 11 | 8 | 3 | 0,73 | Effectuation |
| S17 | 11 | 7 | 4 | 0,64 | Mixed |
| S18 | 12 | 9 | 3 | 0,75 | Effectuation |
| S19 | 23 | 16 | 7 | 0,7 | Effectuation |
| S20 | 26 | 13 | 13 | 0,5 | Mixed |
| S21 | 11 | 10 | 1 | 0,91 | Effectuation |
| S22 | 11 | 9 | 2 | 0,82 | Effectuation |
| S23 | 13 | 11 | 2 | 0,85 | Effectuation |
| S24 | 14 | 6 | 8 | 0,43 | Mixed |
| S25 | 31 | 21 | 10 | 0,68 | Mixed |
| S26 | 13 | 13 | 1 | 0,92 | Effectuation |
| S27 | 6 | 5 | 1 | 0,83 | Effectuation |
| S28 | 33 | 27 | 6 | 0,82 | Effectuation |
| S29 | 31 | 25 | 6 | 0,81 | Effectuation |
| S30 | 23 | 18 | 5 | 0,78 | Effectuation |
| S31 | 25 | 19 | 6 | 0,76 | Effectuation |
| S32 | 18 | 13 | 5 | 0,72 | Effectuation |
| S33 | 13 | 7 | 6 | 0,54 | Mixed |
| S34 | 6 | 5 | 1 | 0,83 | Effectuation |
| S35 | 16 | 13 | 3 | 0,81 | Effectuation |
| S36 | 15 | 5 | 10 | 0,33 | Mixed |
| S37 | 18 | 8 | 10 | 0,44 | Mixed |
| S38 | 14 | 13 | 1 | 0,93 | Effectuation |
| S39 | 21 | 18 | 3 | 0,86 | Effectuation |
| S40 | 27 | 16 | 11 | 0,59 | Mixed |

The EI value for each case is given in Table 8. The higher effectuation index value is, the more dominant effectual activities are found. Table 9 describes the mean EI value for startups regarding their locations (international or Nordic startups), stages (pre, startup, or post phase), and application domain sectors. Pearson chi-squared test shows no significant difference in the distribution of EI values across these categories.

Table 9: EI values among startups in different locations, stages and industry domains

| **Context factors** | **N** | **No Effectuation per case** | **No Causation Per case** | **Mean EI** | **Chi square test** |
|---|---|---|---|---|---|
| **Locations** | | | | | |
| International | 8 | 10 | 3.38 | 0.76 | p-value= 0.80 |
| Nordic | 32 | 11.56 | 4.84 | 0.72 | |
| **Stages** | | | | | |
| Pre-startup | 14 | 12 | 4.64 | 0.74 | p-value= 0.43 |
| Startup | 22 | 11.41 | 4.04 | 0.75 | |
| Post-startup | 4 | 7.75 | 7 | 0.51 | |
| **Industry domains** | | | | | |
| Agriculture, Forestry, Fishing and Hunting | 3 | 9 | 1.33 | 0.87 | p-value= 0.12 |
| Arts, Entertainment and Recreation | 1 | 7 | 1 | 0.87 | |
| Construction | 1 | 14 | 5 | 0.74 | |
| Education | 4 | 7.75 | 6.5 | 0.60 | |
| Health Care and Social Assistance | 7 | 14.86 | 5.29 | 0.71 | |
| Manufacturing | 2 | 16 | 3.5 | 0.84 | |
| Professional, Scientific, and Technical Services | 18 | 11.33 | 4.33 | 0.73 | |
| Transportation and Warehousing | 2 | 11.5 | 8 | 0.57 | |
| Utilities | 2 | 10.5 | 4 | 0.72 | |

## 4. Results and Analysis

The sections below present the data obtained during the study and how this data answers the research questions.

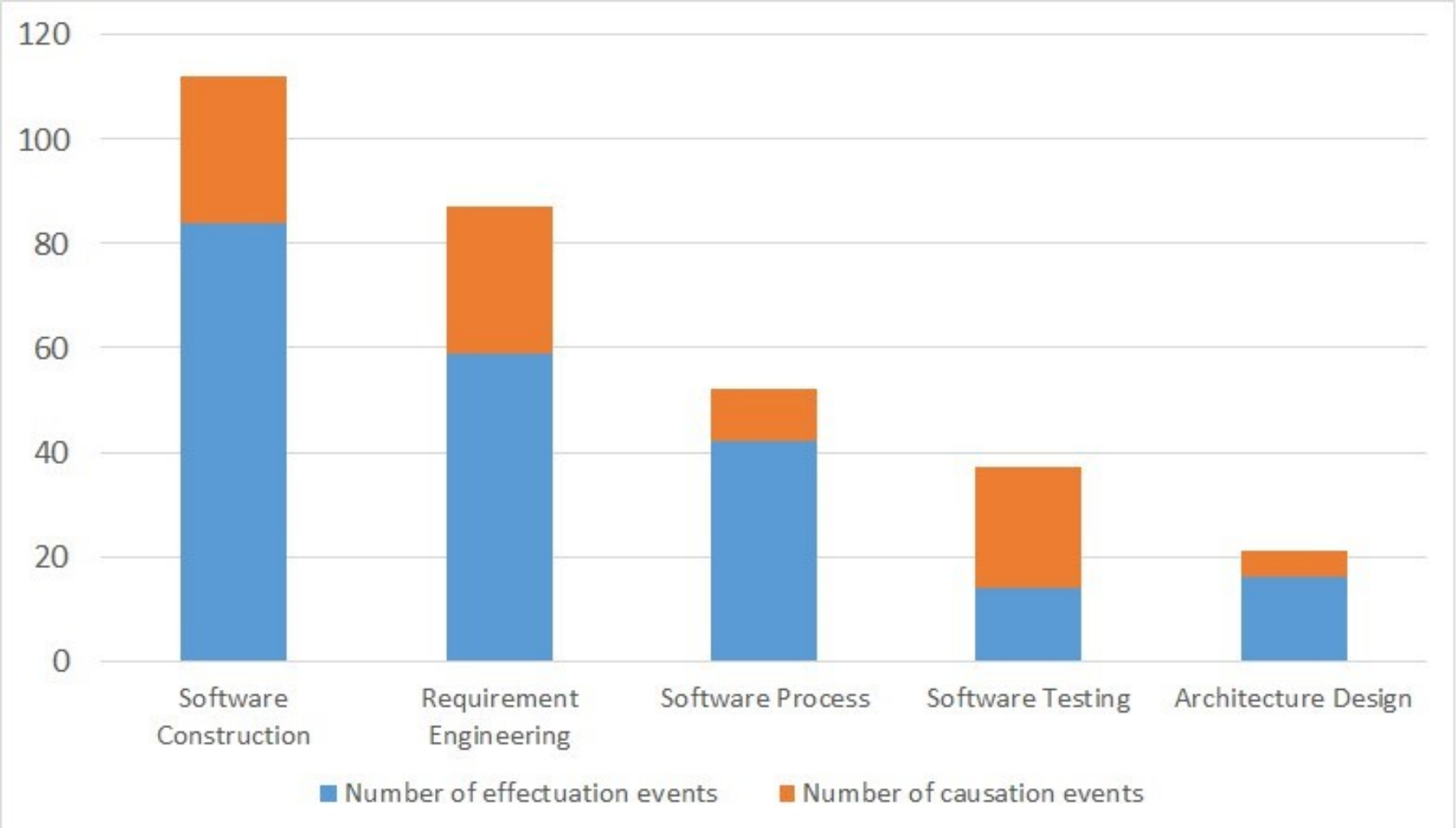

Figure 7 The occurrences of entrepreneurial logics across SE area of activities

### 4.1. RQ1: How do entrepreneurial logics apply to SE activities in startups?

We identified many SE events tagged with entrepreneurial logic. Figure 7 presents the distribution of these events across SE knowledge areas and logic types. We observed the largest number of entrepreneurial events associated with software constructions, followed by requirement engineering, software process, software testing, architecture design, and software maintenance (details are given in Table 10). As our interviewees did not describe their SE activities to the same extent (some focused more on requirement engineering aspects, some talked more about their software processes), the numbers do not represent close relationships among the SE knowledge areas. However, we can see that both effectuation and causation logic occur in all SE knowledge areas. Effectuation logic is the dominant logic in software construction, requirement engineering, architecture design, software maintenance, and software. Testing is the only type of activity where our cases reported that the number of causation events was larger than the number of effectuation-driven events. We present detailed observations in the following sub-sections.

Table 10: Entrepreneurial logics across SE activities

| Area | No. SE Events | No. Effectuation Events | No. Causation Events |
|---|---|---|---|
| Software Construction | 112 | 84 | 28 |
| Requirement Engineering | 87 | 59 | 28 |
| Software Process | 52 | 42 | 10 |
| Software Testing | 37 | 14 | 23 |
| Architecture Design | 21 | 16 | 5 |
| Software Maintenance | 15 | 9 | 6 |

#### *4.1.1. Requirements Engineering*

The thematic codes for entrepreneurial logics in Requirement Engineering are given in Table 11. Requirement engineering activities include the elicitation, analysis, validation, documentation, and scoping of software requirements (Bourque and Fairley, 2014). In many software startups, requirements based on hypothesized business or market demands and the process of requirement elicitation were directly associated with the customer development journey (Blank, 2013; Melegati et al., 2019). Requirement elicitation and management in many software startups can be characterized by effectual logic. The primary sources of requirements are internal stakeholders (i.e., the startup founders) and external stakeholders who can be reached with the existing resources, i.e., the entrepreneurs' personal and professional networks. It is common for startup founders to generate product ideas themselves or based on internal knowledge and then derive concrete requirements by engaging customers early in the development process. This differs from the situation in established companies, where several requirements originate from paying customers or marketing departments.

> We actively looked for the right requirements from our customers. We occasionally searched for any product area. Recently this year, we started a new product feature, and this disrupted one of our old features.

> That decision was made in collaboration with our customers and was implemented with their sponsorship. It is a collaboration agreement in which they finance parts of the development (S14).

In market-driven startups, requirement engineering is often centralized around lead users (von Hippel, 1986), who can express demands currently unknown to the public. They are not the source of requirements, features, and main factors driving startup learning. They contribute to product brainstorming, testing, and feedback or even participate in developing and co-creating new products or services. We found that startups often collaborate closely with some lead users or include them as the startup teams' internal members.

> We sketched the designs of all the ideas we had. We discussed them one by one. The most convincing idea was selected for further development. . . I know the food delivery and stuff like that pretty well, so I am often the winner (S04).

We see that startups might not find many more requirements for their products than what can be collected from their lead users. Startups might start with a sub-optimal set of requirements due to the limited inclusion of lead users (Nguyen-Duc et al., 2017), and the requirement list gradually evolves due to changes in startups' resources and networking positions. It is also possible that not all users' inputs manifest as valuable for customers or a market. Startups often have a low threshold for stakeholder participation and influence on their businesses and products. Depending on internal social capital, startups might involve lead users to different extents. The low threshold seems to be a challenge for startups to identify valuable inputs, and hence, there could be a long journey to identify the winning features.

Moreover, startups often build prototypes as a means of communicating their requirement engineering activities. However, prototype features, such as levels of fidelity and types of prototypes, depend highly on the currently available resources for making such prototypes. For instance, the CEO of S04 mentioned using screenshots that she created in communications with her first customers:

> We worked directly with a customer's organization and learned their current solutions. We described our approach using prototypes like screenshots. It would be hard for them to realize the benefit without concrete examples... (S04).

The prioritization of requirements that match the current development resources is more evidence of effectual logic. Startups might have many innovative ideas, but they will often produce low-fidelity prototypes due to the lack of development resources. Functioning prototypes (MVPs) are often built based on available software components, libraries, frameworks, and even other software products. For instance, many pre-existing libraries are used to develop AI-based software solutions (S21, S32), create new operating systems (S10, S14), or assemble web-user interfaces (S01–S04, S06–S09, S30).

Table 11 Entrepreneurial logics in requirement engineering activities

| Logics | Codes | Explanations |
|---|---|---|
| Effectuation | Engaging in stakeholders within founders' networks | Startups utilize their existing resources and connections to identify, gather, and validate their product requirements |
| | Quantity and quality of requirements depend on lead users involved | Startups actively seek lead users (who deal intensively with the requirements where there is no suitable solution existing on the market) within their resources and capacity |
| | Significance of internal source of requirements | Requirements might significantly come from internal stakeholders |
| | MVPs for requirement communication | low-fidelity MVPs due to the available re-source and time in the startup team |
| | Resource-driven requirement prioritization | Requirements might be adjusted and prioritized due to available technical resources, infrastructures, and code |
| | Tolerance of sudden changes in requirements | Startups tend to accept changing requirements from key customers that might lead to business-driven pivots |
| | Adaptive approaches to prioritize the requirement | Requirement selection and prioritization depend on the business context, iteratively leading to a family of product line |
| Causation | Requirements extracted from a comparative analysis | Product features are identified via successful experience or competitors' products |
| | Plan-driven analysis of markets, customers, and competitors | Startups are not able to develop all collected requirements, and a formal prioritization process is often needed |
| | Avoiding changes to core business value development | Startups often negotiate with customers on requirements that are not aligned with their core business values. Negotiation is often done based on a long-term goal analysis |
| | Dropping new requirements to avoid unforeseen development | Requirements or new features that lead to an unforeseen cost-benefit situation are canceled |

Regarding requirement change management, software companies in the early stages (pre-startup or startup) tend to accept and incorporate significant changes in their feature list, leading to a pivot. This can be explained by the effectual logic attitude towards unexpected events.

> It's very difficult to say no when "giant" customers tell you we need that functionality. If you're going to have us as customers, you will have to make it. We need it in the contract that you have to make it. We also built it [the software product], and we built it bigger and bigger (S11).

The change can also lead to reworking during product development, which startups will need to cope with:

> There will always be new requirements arriving, that is one thing. Sometimes the new requirements disrupt the old ones. At the moment, we are working to disrupt the whole old product, reinvent them, and throw away the whole codebase (S14).

Startups often adopt adaptive approaches to deal with changes in requirements. Requirement selection and prioritization depend on the business context and, iteratively, lead to a product line family. To sum up, many requirement engineering activities are recorded in the association to effectual logic.

Software startups also express causal logic when dealing with requirement engineering. Goal-oriented requirement engineering occurs when defining and planning user stories with limited uncertainties. The analysis of customers, markets, and competitors is goal-oriented and follows some kind of predefined plan. Requirements are then turned into short-term sprint backlogs and are often implemented according to the sprint plans.

> We found out that in Norway, the public only knows one ticketing provider called Company A, which is owned by Company B, which is owned by Company C, a big international company. Then we saw a market possibility for providing a ticketing system, a DIY ticketing system for the smaller venues because Company B and Company C, everything you had to do you did by email correspondence (S11).

When faced with an unexpected change—for example, customers proposing an innovative but peripheral requirement—many startups implement a causation-centric strategy by avoiding unforeseen consequences in developing a product or business, even though this would lead to some customers' requests being adjusted or even dropped.

> We turned down clients that did not believe the [00:17:48] requirement. We make a very clear statement to what we think the future of journalism is, then we pursue that, and the cost of that is neglecting parts of our market (S14).

Observation 1: Requirement elicitation, negotiation, and management tend to be effectuation-driven processes in which startups explore new types of products or customers. Certain finely detailed activities, such as requirement breakdown, estimation, analysis, and validation, tend to be causation-driven when requirements are known to some extent.

#### *4.1.2. Software Construction*

The thematic codes for entrepreneurial logics in Software Construction are given in Table 12. Software construction is the creation of working software through a combination of configuration, coding, unit testing, and debugging (Bourque and Fairley, 2014). In software startups, construction activities apply to both final software products and MVPs. Concerning effectual logic, MVPs are typically developed according to this model in terms of how the speed of implementation, the functionalities, and quality of MVPs mainly rely on the available technical competence in the startup companies. A startup can launch a good front-end prototype very quickly with a user experience expert in the team. A startup can also start with low-fidelity wireframes created by a startup founder who does not have technical competence. The founders of the startups examined in this study created various MVPs, including paper sketches, mockup design tools, and competitor products (S2, S09, S11, S13).

> The first version is like a hack; it took a lot of time to make it up and running. It was impossible for teachers to use because it needed a developer to set up all the network things. It was done in a really hacking manner. Also, it was one instance so it could run one quiz at a time… It failed completely as we just had to throw away the prototype (S09).

The underlying logic is to accept development waste and focus on learning from throw-away prototypes. The minimum effort could also become a wasted effort, for example, when the prototype simulates but does not illustrate. The CEO of S11 introduced the concept of faking a product: "fake it until you make it." Without technical capacities, he demonstrated his business vision with a "faked product," which implies a lack of primary quality, both in terms of functionality and user experience. The CEO of S02 expressed that what they built, in the beginning, is a minimum potential prototype, but not the MVP:

> Building a prototype is like building a fake house. The exterior design is done, you can see how it feels, but the internal part is empty. It helps you figure out if this kind of house you want to live in… We are creating a minimum viable product, not a completely viable one (S02).

Consequently, road mapping and planning for MVPs are often overlooked. After completing an MVP, the creation of the next one is often decided opportunistically. Essential elements of a plan are often neglected: for example, how many MVPs are needed, for what the next MVP will be used, and the criteria for evaluating MVP learning outcomes. Startups accept failure when building MVPs and embrace exploratory development at the cost of economic sacrifices.

Startups are also known to adopt workaround solutions, such as a set of files that is reasonably functional. Developing and testing MVPs at fast rates enables startups to validate their assumptions about their business viability. However, the focus on development speed can also lead to minimum viability. A workaround solution is different from a planned temporary solution, i.e., a piecemeal MVP (Ries, 2011). In some cases, startups have to throw away an MVP that was not designed for long-term use. Startups often aim to develop software early by incrementally adding features into the prototype (Nguyen-Duc et al., 2017). In this way, TD is created as startups focus on speed and neglect quality (Giardino et al., 2016). More seriously, some software products were built using architecture not designed for scale. S09 developed an in-class quiz-based application to check student understanding of lectures in real time; multiple prototypes developed by the CEO or as a student project experimented with different classes. The final prototype, which captured refined design and business ideas, was further developed into a version suitable for launch; however, the release's quality did not match the performance demanded by unexpected growth in users.

In another case, S27 rapidly developed an initial MVP with a hastily built front-end and a hacking back-end function with no security that just achieved minimum performance. The MVP was thrown away, and the company acquired its first seed investment for serious prototype development. Somehow, the launched product contains many components from previous MVPs with a large amount of technical debts. The team deployed the product to customers and extended it further. This later was perceived as a mistake that costed the company significantly:

> At one stage, you just had to drop everything but keep the concept and create it from scratch. The concept was good, the implementation was not bad, but it didn't fit into the commercial world. And at another stage, we needed to get new people with some new minds that could think slightly differently (S27).

Many startups hired external resources, such as local contractors or offshore software vendors, based on their experience and networks. This practice is often the case with non-technical founders or companies with limited in-house technical competence. Examples of local contractors are consultant companies, makers' spaces, student projects, and freelancers. In case S15, skilled contractors were hired to achieve a quick start with a functional MVP. As mentioned by the CEO, the use of external resources enables speedy product experimentation and development. Furthermore, as contractors are not an integral part of the startup, their relatively easy dismissal facilitates scaling-down activities that may be necessary if the startup lacks funding or changes directions. Some startups use local contractors, while others hire offshore vendors. Making use of local vendors can be a feasible option:

> The local one [vendor] delivered very quickly. It is critical that the component from China comes on time, especially when we needed to demonstrate in the week after in the UK. It is always a matter of time. If we could do everything internally, we would have saved a lot of time sending; it would have been great! (S38)

Table 12 Entrepreneurial logics in software construction activities

| Logics | Codes | Explanations |
|---|---|---|
| Effectuation | Resource-based MVP development | MVP development relies on existing and accessible technical competence |
| | Product experiments with tolerance for waste | MVPs are created for demonstration, which is not suitable for long-term use and often thrown away in the later stages of the startup journey |
| | Overlooked product road-mapping and planning | Construction of MVPs usually occurs in an experimental and opportunistic manner |
| | Speed-first MVP development | Startups often need to balance quality and speed to market, and in most cases, time-to-market is prioritized |
| | Recruitment of external competence | Hired developers or contractors are often from the founders' network |
| | Component-based development | MVPs typically contain a significant number of ready-to-use components that can be plug-and-play in a short time |
| | Innovative product development requires exploratory approaches | Innovative products often involve RandD activities that are not purely driven by goals and plans |
| Causation | Short-term plan-based product development | Product development is planned from the requirement to launch, and different efforts are performed to achieve the initial plan |
| | Preventing software constructions from business threats | Software construction might be paused because of financial and business uncertainty |

Most of our startups leveraged existing libraries, frameworks, and components to build a runnable MVP quickly, accessing either paid APIs or Open Source Software (OSS) components. Particularly, the adoption of OSS components was mentioned in all the cases, from using OSS tools (S19) to the integration of OSS code (S02, S03, S05, S20) to participation in the OSS community (S18). The main benefits, including reduced development cost and faster time-to-release, were mentioned by the CTO of (S19) and (S20):

> The things we are doing today, we might not even come to the idea of making it happen if we do not have open-source software (OSS) as an experiment. Without OSS, it would take a lot of time and be very costly (S19).

> It is very hard nowadays not to use OSS artifacts, especially when with Android development (S20).

It was observed that component-based development can influence the product architecture of early MVPs. In S02, OSS JavaScript frameworks were considered the central part of product architecture for the web and mobile applications. It also appears that many advanced technologies were adopted using OSS:

> A core part of our product includes a machine learning (ML) algorithm. We are lucky enough to find ML libraries in C++, and they are entirely OSS (S02).

One possible challenge of using ready-made components is to find a suitable component in terms of maturity, code quality, and level of support, which also appears as an effectuation-driven behavior:

> OSS is used in many architectures and for many purposes... Searching a suitable library was sometimes not so easy but the time was paid back at the end (S05).

The selection process needs to consider functional requirements and quality requirements for the component and the whole product. The CEO of S12 stated that many large companies had offered free APIs to access their data and functionality, integrating them into final software products to consider other issues, such as quality, scalability, and cost. OSS components might not be the minimum available solution, but they reduce the inherent risks of scaling for later phases.

In terms of causal logic, from a short-term perspective, software constructions are plan-based, with concrete expected outputs. Startups also adopt a plan-based product development approach from the requirement elicitation activities to product deployment. So long as the product requirements are identified (e.g., an established sprint backlog), it is expected that the sprint will be operated without many changes.

> For us, this was not the major change because the product was ready, and the customer had the need for it. It was a pretty straightforward delivery for us (S14).

Business and financial stability appear to be important influencing factors on whether causal logic occurs and underlies the software construction process, as in case S03 with secured initial seed funding:

> For the first two or three years, we have been only a product development-driven company. Everything we did was product development... We have grown a lot in Norway by product development, word of mouth, and customer satisfaction (S03).

Startups also express their causation-driven behaviors by avoiding unexpected events and focusing on internal project activities rather than engaging in external interaction. For instance, S25 had to stop its development activities due to uncertain financial conditions in its early stage.

> The operations... stopped, like, one-and-a-half years ago, when we noticed that we were not capable of raising the risk funding for the development of the required technology (S25).

Observation 2: MVP development is typically an experimental and waste-tolerant process, driven by time-to-market, available competence, and internal incentives. Software construction is often opportunistic, and plan-driven coding activities occur in the short term or later stages of a startup's life cycle.

Table 13 Entrepreneurial logics in software design activities

| Logics | Codes | Explanations |
|---|---|---|
| Effectuation | Early customer involvement in solution design | Startup actively involves customers in the design space to co-create business value |
| | Solution design as an experimental process | Solutions for a given customer or market are iteratively visualized through experimental activities |
| | An adaptive approach for a configurable product design | The product design in some cases needs to be adaptable to different customers' requirements |
| Causation | Technical architecture as an optimizable task | Architectural decisions are made with a thorough consideration of cost-benefit trade-offs |

#### *4.1.3. Software Design*

The thematic codes for entrepreneurial logics in Software Design are given in Table 13. Software design represents the problem-solving space where actual business value is planned to be implemented. Software design can include both user interface design and architectural design (Bourque and Fairley, 2014). The effectual logic is apparent in the solution design process, i.e., identifying the best solution for a current customer or market demands. This process might be experimental, means-driven, and change-prone. The process often involves early customers. In S19, the startup not only exposed their MVPs to potential customers quite early, but also used MVPs to involve the customers to their design process:

> Yes, I think it is important to get the customer involvement in the [product] design... Otherwise, it would be a bit scary to launch a new system with assumptions that someone would use it (S19).

Regarding architectural issues as part of the experimental process, S37 said that:

> We have struggled with the choice of platform for the autopilot. Controllers need to be implemented on something, so we have spent a lot of time on embedded components to get the right protocols to control the reserves and get to know what ran on the OSS stuff (S37).

Because of active customer involvement, in some cases, architecture needs to be adapted to cover different requirements:

> Because there are many other parties involved. And many other systems where the interfaces might not be so able to integrate if they are old legacy systems. So they are usually the biggest challenge (S31).

We also observed a customized software design when different requirements arose in the design phase. The architecture for a single product might need to evolve into a product-line architecture with extendibility.

In terms of causal logic, technically speaking, software design is a plan-driven activity. The integration of complexity and other quality attributes in functional software is an achievable task. Architectural decisions are made with a thorough consideration of cost-benefit trade-offs, especially when a system needs some quality attributes for the long run, i.e., performance, and availability.

Observation 3: From a business perspective, software designing is an effectuation-driven process; from a technical perspective, software designing could be plan-based and optimizable.

#### *4.1.4. Software Testing*

The thematic codes for entrepreneurial logics in Software Testing are given in Table 14. Software testing includes four levels: unit, integration, system, and acceptance testing (Bourque and Fairley, 2014). Regardless of levels of testing, we focused on the testing activities that evolve potential or actual users. Product testing is based on assumptions and hypotheses set by the startup about generated value for users and customers. In this sense, product testing is an important mechanism to validate the product/market fit. Many startups do not talk about their testing process in detail. From our observations, startups appear to have more causation-driven testing activities than effectuation-driven ones, as described below.

Effectual logic appears in software testing as the minimum viable testing concept. As found in previous studies (Giardino et al., 2016), software startups prioritize time-to-market over acceptable product quality. This practice is represented by the lack of proper test plans and insufficient testing at different levels. Startups use existing developer resources, such as spare development time, in their milestone-driven plans for testing.

> We prefer to work quickly, and writing tests could double the development time... If these parts are built to be replaced later, then we think there's no point in spending time on testing (S2).

They can tolerate possible losses due to the lack of quality focus. The available resources and equipment then influence the testing activities:

> We are working with a partner to put in place some equipment for further testing, but until now, we have focused primarily on approximate measurements due to a lack of premises and equipment (S39).

When releasing and testing a product version to early adopters, a company may sell the product to others through word of mouth. This fits with the effectual logic concept of "initial customers as partners and vice versa." Overall, it seems that startups with a broad base of potential customers and investors interested in what they are doing as they develop a new product or service have an advantage over those entrepreneurs operating in isolation.

Causal logic is more apparent in the development of particular types of products (e.g., hardware-relevant products) or specific application domains (e.g., automotive, and healthcare industry), where quality is intrinsic to a released software system. In the mindset of these startup founders, testing is as essential as implementation. High quality in hardware development is vital because of the cost associated with production and quality mistakes, which dramatically affect the perceived functionality of the product (Berg et al. 2020). In contrast to software products, it is challenging to implement changes and improve the quality after the product has been produced and assembled.

> Failures can cause high costs, more work, and, at worst, a security issue to make sure no one gets hit if it [a flying drone] falls. This is opposed to a car or a boat because testing them is much easier. Setting up robust tests and making a foundation for testing for something that can fail in the air is a unique challenge with flying things. The quality has to be better, and it is not easy to test things... We must accept that the fastest way isn't always the best one. For flying, it is important to do things properly instead of choosing quick solutions (S37).

Some companies decided to do test-driven development by developing both requirement description and test cases, using the test cases to track the software development, as illustrated in case S40:

> Test-driven development is also changing now a little towards acceptance of test-driven development. So, we can write tests that customers can also read and verify by themselves that they are passing and that we are implementing the right features. Also, we are moving more towards automated end-to-end tests, that the test begins from the user interface and ends... (S40).

Observation 4: While system testing and user acceptance testing are often results of causation-driven behaviors, effectuation-driven testing is often applicable for demonstration.

Table 14 Entrepreneurial logics in software testing activities

| Logics | Codes | Explanations |
|---|---|---|
| Effectuation | Minimal viable testing | Startups might perform testing just enough for purposes of demonstration or launching |
| Causation | Testing is by-designed in specific types of products | Hardware-related products often require heavy upfront testing |
| | User acceptance of test-driven development | Startups with quality as value proposition need to achieve their plans for user acceptance tests |
| | Test-driven development | Test plan are often made at the same time with requirement specifications in hardware-related products |

#### *4.1.5. Software Maintenance*

The thematic codes for entrepreneurial logics in Software Maintenance are given in Table 15. Software maintenance in SE is about modifying software products after delivery to correct faults and improve performance or other attributes (Bourque and Fairley, 2014). In startups, software maintenance and construction are often mixed when providing running software for some customers and, at the same time, developing new features or new variants of the product. Where effectual logic is concerned, startups often take on many maintenance tasks as they support their first bespoke customers. In these cases, product improvement and new features are not typically planned, and customer satisfaction is an important criterion that directs further development.

> Question: Have you planned for a way to upgrade the software of sold gloves?
>
> Answer: We have not thought about that. We have assumed that if the user thinks there is something wrong, then the user will contact us. Then we help the user with the error that has occurred. We have no analytical overview of the products that are out there (S40).

Table 15 Entrepreneurial logics in software maintenance activities

| Logics | Codes | Explanations |
|---|---|---|
| Effectuation | Customer-driven software maintenance | Maintaining tasks tailored to specific customers |
| | Contingency approach of managing TDs | Depending on contexts, the debts can be managed, accepted, avoided, or ignored |
| | Reacting to bespoke change requests | Maintenance tasks occur from new feature requests or bug fixes for bespoken customers |
| Causation | Scheduled management of tools and infrastructures | Maintenance tasks, including infrastructure and configuration management, are typically scheduled and repeated |
| | Planned software maintenance | Software maintenance tasks are planned along with development and testing |

Another trigger for effectuation-driven software maintenance is the TD incurred during software construction (Giardino et al., 2016). The metaphor implies that "interest" has to be paid during maintenance and development activities and that the "principle" should be repaid, i.e., with code refactoring, at some point for the long-term health of the software product (Krutchen et al., 2012; Seaman and Guo, 2011). While the startups agreed that TD trade-offs are crucial for their businesses, they each handled the debt differently. It is often uncertain whether the impact of "work-around solutions" on later maintenance tasks when products are operating in a customer environment.

> In the beginning, we made a lot of mistakes, but they didn't last long... Now that you started re-coding the system, leaving roughly six months of work behind... you said let's leave it there (S11).

The effectual logic here is shown by the contingency approach that TD can be purposefully avoided, fixed, or ignored.

In terms of causal logic, we also observed that, within these startups, maintenance tasks, including infrastructure and configuration management, are typically scheduled and repeated regularly. Software maintenance can also be planned to some extent to avoid unexpected incidents and optimal in terms of cost-benefits:

> It was implemented in such a way that it was not difficult to work on it or further develop it (S40).

Observation 5: Software maintenance in software startups occurs opportunistically. Dealing with TD is effectuation-driven by nature. Startups often throw away systems that are not working and develop new systems rather than reverse engineering the faulty product.

#### *4.1.6. Software Process*

The thematic codes for entrepreneurial logics in Software Process are given in Table 16. Startups are known to adopt a lightweight and agile workflow rather than following a specific formal method (Pantiuchina et al., 2017; Nguyen-Duc et al., 2017). Many startups do not have actual processes or a systematic way of working because they do not often prepare for a long run. Align with the effectuation approach, software startups tend to be agile or even ad-hoc and reactive:

> Yes, we have always been working in an agile way. We are not adhering to any specific method, but we cannot do long-term specifications. That is not doable in an industry that is changing very rapidly. We have always been working with long-term visions but with short-term specifications. The way we developed specifications, it is always with the collaboration with the clients or the customers (S14).

Agile development was mentioned as the best approach to achieve speed and agility in startups. The CEOs related agility to less upfront planning and the short-term driven evolution of the startups. They also mentioned the speed of prototyping, development, and fast time-to-market when asked about an agile approach. Employees at the startups stated that full control of development activities and partnerships would prepare them to respond to unexpected changes. Some startups also highlighted the importance of team collaboration over defined processes. The adoption of certain agile practices or approaches might differ between the development of hardware and software elements:

> Our MVP is relatively simple, while software changes likely happen all the time. We are still trying to find what is the right way to do it (S37).

Many startups characterized their workflow as a trial-and-error approach, adopted to deal with uncertainty in business and technology. It is worth noting that technological uncertainty might be due to the complexity of technology and the team's available technical competence.

> Typical sprints are anywhere between one and five days, and we always give very small steps to make sure that we don't head down a blind road, a blind alley. To make sure that we all understand what we're doing without making (S12).

Startups might not know which development approaches are practical for them due to their relatively short operation history. This is different from established companies, where they have learned and adopted stable working approaches. The journey of learning about processes and practices is rather means-driven than goal-driven; the processes are proposed by and adopted for the available resource in the startups. The effectiveness of the adopted process is then determined by the current startup team, current resources, customers, and products.

Table 16 Entrepreneurial logics in process management activities

| Logics | Codes | Explanations |
|---|---|---|
| Effectuation | Short-term planning | Short iterations are commonly adopted with a vision of at most six months in advances |
| | Change-prone and dynamic development environment | Startups might face difficulties in adopting a set of specific development approaches due to the quick change of the project context |
| | Self-defined workflow | It is typical for many startups to adopt no formal guided development approaches |
| | Evolving working processes and practices | Changes in organization or product might trigger the need to try better development approaches |
| Causation | Plan-driven adoption of software processes | Startups might pursuit a strategic goal of adopting software processes and practices |

> We have not made any such routines, so we are at that stage that we learn that we should do it. We started very sharply, and we have not yet reached a point where we have realized that it could help us. I know that we could probably have served more formal routines (S39).

Startups often react to their environmental contingencies by adapting their workflow to fit the new financial, organizational, and managerial conditions. Again, the means-driven attitude applies to the startups because their adjusted working approaches will depend on their internal competence and experience with methodologies. A startup team would not be likely to try out a Lean Startup approach if they do not have anyone in a team with prior experience with this style.

> We came to a crossroads in February, where we decided to let one tech team continue to work on it, and one team started to work with flex sensors. We wanted to see if we could get a faster prototype by changing the solution method (S40).

In terms of causal logic, we also observed cases where plan-driven product development was adopted from the beginning. For instance, S03 had little uncertainty about their business due to the investment and precise product requirements. The product development was prioritized and planned in a year based on the formal analysis of product requirements and a stable development team:

> In the next 6 to 12 months, we are going to move into a real strict agile process. Because we have our daily stand-ups and our backlogs and stuff like that, but we try to keep it a little loose (S03).

Observation 6: Software startups are characterized by self-defined, adaptive, and opportunistic workflows. The evolution of practices and processes is expected through startup development.

### 4.2. RQ2: How do entrepreneurial logics apply to software development at the company level?

Insights from RQ1 do not give us a comparative view among startups regarding how they adopt entrepreneurial logic during their product development. We calculated the Effectuation Index (EI) using events extracted from SE activities for each startup case. There were no startups that included only causation logic or effectuation logic. To search for a possible explanation for the application of effectual logic or causation logic, we conducted a Chi-square test (as shown in Section 3.4.4) in Mean EI values across startup locations, phases and industry domains. Qualitatively, we looked at the common codes that are identified as effectuation-driven or causation-driven activities and summarized them at the company level. The type description below applies to a state of a startup, without excluding the possibility that startups shift among these types. Observations from 40 startups showed that at a certain point in time a startup can be characterized as either an effectuation-dominant startup or a mixed-logic startup. The list of startups cases according to their types is given in Table 17.

Table 17 Entrepreneurial logics occurred in the company level

| **Startups type** | **Definition** | **Common conditions** | **Startup Cases** |
|---|---|---|---|
| Effectuation-dominant | Startups that are experiencing the major number of their product development activities under effectuation logics | Great level of uncertainties<br>Limited resources<br>Frequent iterative processes<br>Technical debt<br>Pivot-ready | S01, S03, S04, S05, S07, S08, S10, S11, S12, S15, S16, S18, S19, S21, S22, S23, S26, S27, S28, S29, S30, S31, S32, S34, S35, S38, S39 |
| Mixed | Startups that have significant number of effectuation-driven and causation-driven activities | Reduced uncertainties<br>Acquired team competence<br>Agile-like procceses<br>Process improvement | S02, S06, S09, S13, S14, S17, S20, S24, S25, S33, S36, S37, S40 |
| Causation-dominant | Startups that are experiencing the major number of their product development activities under causation logics | Managable level of uncertainties<br>Traditional software development procceses | None |

#### *4.2.1. Startup Type 1: Effectuation-dominant*

Effectuation-dominant startups (27 out of 40 cases) often initiate with some unique advantages: for instance, a product idea that did not previously exist, a market segment with little competition, a group of talented developers, or an existing source of customers. These startups strongly emphasize personal knowledge as the starting point, i.e., the founders are competitive in business competence or technical competence. They also rely heavily on their internal resources. This can be seen from startups deriving the product requirement internally or with existing requirements (Klotins, Unterkalmsteiner, Gorschek, et al., 2019). The requirement elicitation process might involve internal or external (or both types) stakeholders (Melegati et al., 2019). These stakeholders include both lead users and individuals who commit to resources, and in many cases, startups need to discover the most valuable requirements. Social capital, the relationships between people in various networks, is critical for startups because these companies often explore their network to identify new requirements, explore business opportunities, and recruit partners or employees.

Effectuation-dominant startups also rely only on resources that they are willing to lose. Almost all of our effectuation-driven startups threw away many MVPs, even high-fidelity ones (Duv and Abrahamsson, 2016). Startups are willing to take the risk that their products or features are not desired in the market and, in most cases, are ready to pivot to a new idea if necessary (Sarasvathy, 2001). In these startups, technical redirection can happen any time new information is unearthed. These startups do not focus heavily on components and system testing. Software maintenance is carried on in parallel with development with a short-term focus on current customer satisfaction. From this perspective, phenomena such as TD or speed over quality are probably unavoidable and perhaps a way for startups to prepare for the possible losses incurred with a technical pivot. Effectuation is an iterative process, and startups learn continuously from their experiences. We also found that the learning process is *ad hoc* and informal, lacking knowledge discovery, extraction, and storage. This generates challenges in the future when startups scale up their products and organizations.

Observation 7: An effectuation-dominant startup focuses heavily on internal resources and social capital. The startup embraces a focus on speed over quality, neglects quality assurance investment, accepts TD, and tolerates technical pivots to deal with uncertainty.

#### *4.2.2. Startup Type 2: Mixed-logic*

Compared to effectuation-dominant startups, mixed-logic startups (13 out of 40 cases) have reduced uncertainties regarding their markets, funding, and team conditions. Either they are spin-offs from established companies and inherit well-defined problems with existing customer contacts (S06, S20), they have evolved into more stable stages (S09), or they operate in a regulated domain (S40). In these startups, we found several middle-term and long-term plans regarding their product development. These startups emphasize their plan-based analysis, selection, and prioritization of requirements to a (partly) validated market. When they implement their goals, the startups tend to ignore external influences, opportunities, and requirements and instead focus on achieving their visions. Product development is more similar to methodologies reported by established companies. Software development methodologies become more of a concern when startups look for productivity, quality, and a sustainable working experience. In these companies, the significant difference is the investment in testing activities. We also observed the adoption of formal approaches with a large upfront sum, such as test-driven development (S40). Activities such as architectural designs and software maintenance contain a lot of planning and analysis.

Observation 8: A traditional SE process and practice is relevant to a mixed-logic startup where both business and product development is relatively and subjectively predictable. A mixed-logic startup is often the continuation of an existing business or a startup at a certain maturity level.

#### *4.2.3. Startup Type 3: Causation-dominant*

Startups that are experiencing the major number of their product development activities under causation logics are not observable from our sample. By definition, these startups would adopt long-term and analytics-driven approaches. The idea type-3 startups might have an analytical approach to customer requirements, stable value propositions which unlikely to change in a short-term perspective, an early and clear overview of their product architecture, adopt principled software development processes and practices including test-driven development, continuous integration, and DevOps (Section 4.1).

## 5. Discussion

### 5.1. Discussing the Primary Observations

By applying entrepreneurial logics, this study explains many findings from previous software startup research. Melegati et al. (2019) reported that requirements engineering has multiple influences and helps explore the market opportunity and devise a feasible solution. We observed that requirement elicitation and negotiation tend to be effectuation-driven. Startup founders use their current knowledge about technology, markets, cultures, and social capital to identify the product's market fit. Klotin, Unterkalmsteiner, Chatzipetrou, et al. (2019) reported a survey result showing that internal sources, such as brainstorming and the invention of requirements, are the most popular requirement sources. The authors also reported challenges establishing contact with their potential customers and involving them in the product work. In our findings, we showed a means-driven principle, and the startup's social capital would likely shape this challenge. We emphasize that the real problem of requirement engineering in startups is that many customers' requirements would probably not possess the right inputs for product-killing features. Effectuation logic helps to explain why a "lean" approach suitable for User-Centered Design in software startups (Hokkanen et al., 2015). Startups in the early stages often search for lead users from their social capitals and this process is probe-and-sense without detailed plans in advance.

Many startups are not successful in learning from their MVPs due to the effectuation-driven behaviors. They tend to overlook product roadmapping and planning, reuse MVPs for many different purposes and in different scenarios (Duc and Abrahamsson, 2016), which leads to accumulated TD (Giardino et al., 2016). With the attitude of tolerating for (prototyping) failure, MVP building process is seen as a waste-tolerant process rather than a validated learning process in startup and that this process seems to be driven by time-to-market, available competence, and internal incentives. To improve the situations, lightweight guidelines at operationalization levels would be a possible approach (Bosch et al., 2013, Nguyen-duc et al., 2020).

Software testing is a particular engineering area with a dominant number of causation-driven events. On the one hand, we have observed the so-called "minimum viable testing" approach as an inappropriate quality assurance approach. On the other hand, startups often mention testing as a plan-driven endeavor. This happens in startups developing quality-critical products (e.g., hardware startups), safety, and security-critical domains (e.g., health-care and automotive), or in a startup with established software development methodology (process improvement). Klotin, Unterkalmsteiner, Chatzipetrou, et al. (2019) argued that startups could benefit from more rigorous testing practices. We can agree with this observation only when startups have quality attributes as their core value propositions. When MVPs are released for events such as demonstrations and funding pitches, testing is effectuation-driven and minimalistic. Giardino et al. (2016) pointed out several contributing factors to accumulated TDs in software startups, including lack of architectural design, automated testing, and minimal project management. These factors fit well with effectual logic as startup members are prepared for changes, associate change with affordable loss, and avoid excessive upfront investment in design, implementation, and testing. Throw-away work seems to be a natural part of the startup, and reuse seems to be opportunistic.

Regarding software processes, software startups can be characterized by their opportunistic workflows. We would argue that the entrepreneurial logic would determine which software procceses and practices are adopted. Pantiuchina et al. (2017) reported that startups adopt agile practices differently and communication practices, such as daily standup meetings, are not common among startup teams. Their findings of the overwhelming of speed-related practices reflect the adoption of effectuation logic in software construction and testing. Klotin, Unterkalmsteiner, Chatzipetrou, et al. (2019) showed that start-ups often have communication issues, shortages of the domain, and engineering expertise. From the view of effectuation logics, startups tend to include any people who can contribute to their value proposition. This can solve the need for competence and knowledge in a short time but poses a challenge of team cohesiveness. While a team often needs to go through several steps to reach their optimal performance (Tuckman, 1965), startups might have a challenge reaching this performing stage.

Our findings propose a classification of startups into either Type One (effectuation-dominant), Type Two (mixed-logic) or Type Three (causation-dominant). Once startups are organized into distinct types, it will be easier to reason about their engineering activity decisions. Traditional SE processes and practices are more relevant to causation-domain startups (Type Three) or mixed-logic startups (Type Two), where both business and product development are relatively and subjectively predictable. This aligns with some observations from previous studies, i.e., startups in the early stages adopt less formal management practices than those in the post-startup stage (Klotin, Unterkalmsteiner, Chatzipetrou, et al., 2019). We argue that a startup classification is a contingency approach, in which both internal factors (e.g., founders' experience and characteristics, team competence, and available technologies) and external factors (e.g., institutional factors, startup ecosystems, and market conditions) might lead to startups becoming one of the three types. The cross-sectional view on startup classification does not mean a consistent occurrence of the logics in all engineering activities, e.g. causal logics can still be observed from effectuation-dominant startups. Hence, we suggest the reasoning for startup tactics should be done in the association with SE decisions to be taken and its contexts. Besides, we do not capture the dynamic aspect of startup types, i.e. startups might be in a transition from Type One to Type Two or from Type Two to Type Three. A study of logic shifts, as shown in entrepreneurship research (Reymen et al., 2017), needs to be done in a longitudinal manner.

### 5.2. The Applicability of Entrepreneurial Logics in Software Startup Engineering

Exploring and comparing causation and effectuation logic to make sense of startups' business activities are widespread in business research (Sarasvathy, 2001; Sarasvathy and Dew, 2005a; Chandler et al., 2011; Reymen et al., 2015; Reymen et al., 2017; Harms and Schiele, 2012; Smolka et al., 2016). Sarasvathy (2001) emphasizes that causation logic is more suitable for existing markets, and effectuation logic is more suitable for new markets and products. Chandler et al. (2011) illustrate the occurrence of these logics in business experimentation. Flexibility regarding unforeseeable events in effectuation has been contrasted with carrying out a planned strategy under causal logic (Reymen et al., 2015). Our observations show that both kinds of logic can be found in different SE areas of activities. After comparing our results with those of previous studies (Giardino et al., 2016; Melegati et al., 2019; Klotins et al., 2019; Tripathi et al., 2019), we understand the logics behind how some engineering activities are carried on. Indeed, after quantifying the number of logic-driven events across 40 startups, we did not observe any case adhering to only one logic. We do not have enough insight into each case to conclude possible patterns of adopting logics, engineering activities, and their consequences; however, effectuation logic appears to be the primary principle behind product development in the early stages of startups. Moreover, as SE deals with a systematic development approach, we would expect more plan-based and analysis-driven activities in software startups that invest in their workflow. Mansoori and Lackéus (2019) propose a framework for applying entrepreneurial methods consisting of the three levels of logic, model, and tactics. While we have not seen the consistent appearance of causation and effectuation at the model level, we observe that they are relevant at the tactical level.

Tactics connect the abstract nature of the logics to the tangible realm of practices. Tactics are often detailed and specify the context of use and the outcomes of action. Several engineering phenomena, such as MVP, TD, lead users, and test-driven development, can be described using causation or effectuation logic at the tactic level. Reymen et al. (2017) studied decision-making logics in four high-tech startups. The authors found both effectual and causal logics in different parts of startups' business models. In our own study, we see the appearance of both logics in different parts of the product development lifecycle. Harms and Schiele (2012) found that, for business development, the entrepreneurs' experiences might also influence the choice of entrepreneurial logic, not only the surrounding environments. Dew et al. (2009) also found that entrepreneurial experts frame decisions using effectual logic while novices use a causation-driven approach and tend "to go by the textbook." (Dew et al. 2009, p. 1) It might also be that the dominant decision-making logic may shift several times (Reymen et al., 2015), and both decision-making logics may co-exist according to the different degrees of uncertainty in the market and technology or the number of decision-makers involved (Nummela et al., 2014). We observed some contextual variables that might determine whether a effectual mindset or a causal mindset should be in place, including business uncertainty, startup maturity, funding situations, and expertise in engineering methods regarding SE activities. Future work can further investigate these as factors that precede the choice of logic in software startup engineering.

Existing research has also revealed that it is possible to observe both kinds of entrepreneurial logic in different stages of a startup. Smolka et al. reported that causation and effectuation both have positive effects on business development (Smolka et al., 2016). Founders who are resource-driven and engage in planning activities tend to have better startup performance. This is a fascinating observation; however, we did not have enough insight to validate this combined effect in software development contexts. The logic shifts may also explain the changes in startups' business strategies, marketing approaches, and workflows (Reymen et al., 2017; Sarasvathy, 2001; Harms and Schiele, 2012). Looking at startups in different stages, we hypothesize that startup founders can shift their logic from effectuation-driven towards causation-driven by gradually establishing their workflow in different SE activities.

### 5.3. Threats to Validity

In qualitative research, scientific validity must be addressed to replicate research and ensure that the findings are trustworthy (Yin, 2003; Runeson and Höst, 2009; Cruzes and Dyba, 2011). To ensure the validity of this study, we followed the validity guidelines from Runeson (Runeson and Höst, 2009). Construct validity ensures that the studied operational variables represent the construct we aim to investigate according to the research questions. In our study, the components were developed based on existing literature (McKelvie et al., 2020). This study's measure of entrepreneurial logic is based on approaches reported in previous studies (Reymen et al., 2015; McKelvie et al., 2020). Our interview questions reveal major key events in each startup, reflected by CEOs or startup co-founders with insights into business and product aspects. A possible risk here is bias in data, i.e. interviewee might focus on the most improvised or messy aspects of the product development. We implemented some measures against this threat. First, we read again the interview transcripts, which captured also emotional expressions to recall how the interviewee presented themselves. Second, we investigated the context of main labeled events to detect if there is any visible bias. It is not easy to understand a startup and its cultural, institutional, and contextual factors within a single interview of about 60 minutes; therefore, we compensated by collecting data about the startups through incubator and company websites before interviews. We also talked to co-authors or researchers (if available) who have connections to the case to better understand the socio-cultural context of the case.

To improve the study's reliability, we invited all participating startups to proofread the (part of) results to ensure their conformance with reality. Moreover, we had several rounds of discussions after data analysis among authors to allow alternative interpretations and regulate possible over-interpretations. The review process from the Empirical Software Engineering journal also helps us critically reviewing initial research questions and make additional adjustments and analyses.

Internal validity concerns causal relations between investigating factors, such as our entrepreneurial logic, and making engineering decisions. Our study explores the logic's occurrence across types of SE activities and startup cases and does not aim to associate a relationship. Therefore, this particular limitation is not a concern in this study.

External validity refers to the extent to which the findings are generalizable beyond the context studied. For qualitative studies, the intention is to enable analytical generalization where the results are extended to companies with common characteristics. We have tried in different ways to achieve the diversity and representative of our cases. However, it was very difficult to reach out and talk to startups. We did not detect any systematic bias due to the possible differences between the ones who accepted and the ones who refused to participate. By emailing and talking via phone, professional networks and nearby startups have a slightly lower turnover rate than startups from Crunchbase. However, when approaching startups via personal introduction or meet in person, there is a significantly higher chance to acquire their participations.

Our case sample is skewed toward specific geographical locations (Norway and Nordic countries in general), startup phases (dominated by companies in either pre-startup or startup stages), team sizes (mostly between three to 20 people), and funding model (mostly by bootstrapping). Consequently, it would be safe to relate these findings to startups with similar characteristics (i.e., European software startups). Startups from other American countries or startups already in a growth stage might not share the features observed in the majority of our cases; they may be more causal than effectual, for example. Another remark is that our findings apply to the startups at the time at which they were investigated. The study was not designed as a longitudinal case study; hence, we do not claim that entrepreneurial logic will appear in the same way in these startups at another point in time.

Reliability refers to the extent to which data and the analysis are dependent on the specific researchers. We have defined and validated interview protocols with colleagues. Some interviews were in Norwegian. We tried our bset to preserve the actual meaning of respondents via the transcription. Recordings were transcribed shortly after each interview to mitigate bias. We have cross-checked the analysis results between the first and second authors of this study, and a high consensus level was reached. Additionally, we compared findings to related literature (Giardino et al., 2016; Hevnera and Malgonde, 2019; Klotins et al., 2019; Melegati et al., 2019), examining similarities, contrasts, and explanations. Such comparisons have enhanced the internal validity and quality of our findings (Eisenhart, 1989).

## 6. Conclusions

As software startups find themselves operating in uncertain, risky, and dynamic environments, existing software development approaches have limited applicabilities due to their prediction-based theoretical underpinnings. Our goal is to increase current knowledge about SE in startup contexts by adopting the entrepreneurial logic lens. From a qualitative survey of 40 startups, we observe dominant effectuation-driven software development behaviors that focus on requirement engineering, software construction, process management, software design, and maintenance. Effectuation-driven approaches promise to develop different processes, models, and tactics that welcome uncertainty and risk. We showed that both entrepreneurial logics occur and help advance the current understanding of the *how* and *why* of engineering processes and practices in startups. For instance, TD acceptance, requirement identification, and MVPs tend to be driven by effectual logics, while causal logic drives test-driven development.

Future research will explore how best to build software development methods that incorporate aspects of entrepreneurial behavior logic. We propose three potential areas of future research:

- Making the right decisions is essential for entrepreneurial application success in software startups. The effectuation-driven approach to software development supports a new way to take actions that fit existing means but still consider the long-term goals.
- We need to better understand the influence of entrepreneurial contexts on the occurrence of behavior logics. We do not have enough data to compare and analyze different environment conditions and relate them to the frequency of entrepreneurial logics across SE activities.
- Logic can shift from effectuation-driven to causation-driven software development. We had mostly cross-sectional views into startup cases, which limited our observation of the behavior logics' temporal evolution. Founders and managers need to understand how and under which conditions the effectuation-driven behaviors change to causation-driven ones.